\documentclass[twocolumn,aps,prl,showpacs,preprintnumbers, superscriptaddress, amsmath,amssymb]{revtex4}

\usepackage{stmaryrd} 
\usepackage{stmaryrd} 

\DeclareMathOperator\arcsinh{arcsinh}

\newcommand{\micrometer}{\mu\mbox{m}}

\usepackage{amsmath}
\usepackage{amsfonts}

\usepackage{amssymb}
\usepackage{graphicx}%

\begin{document}

\author{Andre G. Campos}
\email{agontijo@princeton.edu}
\affiliation{Department of Chemistry, Princeton University, Princeton, NJ 08544, USA} 

\author{Renan Cabrera}
\email{rcabrera@princeton.edu}
\affiliation{Department of Chemistry, Princeton University, Princeton, NJ 08544, USA} 

\author{Herschel A. Rabitz}
\affiliation{Department of Chemistry, Princeton University, Princeton, NJ 08544, USA}

\author{Denys I. Bondar}
\affiliation{Department of Chemistry, Princeton University, Princeton, NJ 08544, USA}

\title{ Analytic Solutions to Coherent Control of the Dirac Equation }

\date{\today}

\begin{abstract}
A simple framework for Dirac spinors is developed that parametrizes admissible quantum dynamics and 
also analytically constructs electromagnetic fields, obeying Maxwell's equations, which yield a desired evolution. 
In particular, we show how to achieve dispersionless rotation and translation of wave packets. 
Additionally, this formalism can handle 
control interactions beyond electromagnetic. This work reveals unexpected flexibility of the 
Dirac equation for control applications, which may open new prospects for quantum technologies.
\end{abstract}


\maketitle

\emph{Introduction}.
The common aim of quantum control is to find a tailored external electromagnetic field to steer the ensuing 
dynamics in a
desired fashion \cite{brif2010control}. This capability, in particular, is enabling 
quantum technologies with the prospect of revolutionizing metrology, 
information processing, and matter manipulation.
However, little is known about the control of the Dirac equation 
in spite of its modern applications reaching into nearly every domain 
of physics, going far beyond its original intention \cite{greiner2000relativistic,bagrov2014dirac}.
For example, lasers have already reached intensities where
light-matter interactions must be described within the Dirac theory \cite{RevModPhys.84.1177}. 
Studies of the properties of heavy elements led to the establishment of relativistic quantum 
chemistry \cite{grant2007relativistic,autschbach2012perspective,schwerdtfeger2015relativistic,pavsteka2017relativistic}
based on the Dirac equation.
Moreover, there is a growing list of low energy systems emulating  
Dirac fermions in solids \cite{novoselov2005two,katsnelson2006chiral,hasan2010colloquium},
optics \cite{otterbach2009confining,ahrens2015simulation}, cold atoms
\cite{boada2011dirac, suchet2016analog}, trapped ions
\cite{gerritsma2010quantum,blatt2012quantum}, and circuit quantum electrodynamics
\cite{pedernales2013quantum}.

The Dirac equation is commonly expressed as \cite{greiner2000relativistic}
\begin{align}
\label{Dirac-Equation}
    \gamma^{\mu}[ i c \hbar \partial_{\mu} - c e A_{\mu}] \psi = mc^2 \psi,  
\end{align}
where the summation over repeated indices is adopted, $\psi$ is a four-component complex spinor, 
$m$ 
is the mass, $c$ is the speed of light, $\gamma^{\mu}$ are 
the $4\times 4$ so-called gamma matrices, $A_{\mu}$ is the four-vector potential and  $\mu=0,1,2,3$.

The Dirac equation (\ref{Dirac-Equation}) can be viewed as a ``first quantization'' approximation
to QED. The solutions of Eq. (\ref{Dirac-Equation}) exclude effects such as radiation reaction
and particle creation/annihilation prominent at ultrarelativistic energies. 
Nevertheless, Eq. (\ref{Dirac-Equation}) provides a mean-field description 
of relativistic effects at low and moderate energies. 
A moving Dirac electron generates the current 
$J_D^{\mu}  = \psi^{\dagger} \gamma^0 \gamma^{\mu} \psi$ 
that emits secondary radiation, which is not accounted for by Eq. (\ref{Dirac-Equation}). 
Therefore, a solution of the Dirac equation is physical 
if the energy loss due to the secondary radiation is much smaller than the electron kinetic energy.
This criterion should be satisfied in the applications of the Dirac equation to quantum control.

In this Letter we present the framework of \emph{Relativistic Dynamical Inversion} (RDI) opening up
a new route to coherent control for the Dirac dynamics: 
Given a desired wave packet evolution, we \emph{analytically} design electromagnetic control fields 
obeying Maxwell equations.
This should be compared with other techniques  
such as shortcuts to adiabaticity \cite{deffner2015shortcuts,song2016robust,deffnerDelCampo2014ShortCutAdiab}
analytically constructing interactions that often go beyond electromagnetic fields.

The purpose of the current work is to solve the 
following problem: Given an arbitrary (desired) spinorial spacetime wave packet $\psi$, 
find an electromagnetic field  $A_{\mu}$ such that Eq. (\ref{Dirac-Equation}) is satisfied.
This is accomplished by RDI in two steps: First, we verify the attainability of the given evolution 
$\psi$ by assessing the existence of the underlying $A_{\mu}$ leading to valid Maxwell equations; 
second, if it exists, an explicit form of $A_{\mu}$ is obtained.
Moreover, the method can also be used to assess for attainable dynamics.

The task of constructing the control field yielding the desired dynamics at all times 
and positions is one of the most important and challenging problems in quantum control.  
In particular, transporting coherent wave packets without disturbance is a required 
building block in quantum technologies.
RDI allows for finding analytic solutions not feasible by other current methods. 
This is possible due to unique properties of the Dirac equation.

Exact solutions of Eq. (\ref{Dirac-Equation}), a system of four partial differential equations,
are rare. The vast majority of them are for highly symmetric
stationary systems \cite{thaller2013dirac,bagrov2014dirac,eleuch2012analytical}.
Furthermore, finding exact solutions with probability densities having finite integrals over the whole 
three dimensional space is a formidable task. 
Only a handful of solutions for time dependent dynamics exist 
\cite{varro2013new,bialynicki2004particle,Oertel-DiracInversion-2015,kaminer2015self,hayrapetyan2014interaction,Birula-DiracAngularMomentum2017,BarnetRelativisticVortices2017,HeinzlClassicalQuantum2017}.  
Most of the investigations call for either semiclassical methods \cite{lazur2005wkb}
or numerical calculations 
\cite{braun1999numerical,mocken2008fft,bauke2011accelerating,fillion2012numerical,fillion2016galerkin,lv2016numerical,
cabrera2016dirac}. 
In addition to being computationally demanding, commonly used numerical schemes are plagued 
by unphysical artifacts at the fundamental level \cite{hammer2014staggered,hammer2014single}; 
thus, there is a need for  systematic construction of analytic solutions. 
RDI fulfills all these needs by providing stationary 
as well as time-dependent exact solutions integrable in two and three dimensions.

RDI \emph{simultaneously} seeks the state $\psi$ and the vector potential $A_{\mu}$ describing physically admissible dynamics.  
Considering that Eq. (\ref{Dirac-Equation}) is \emph{
bilinear} with respect to both $\psi$ and $A_{\mu}$, it may seem that the proposed approach is even 
more challenging
than solving the \emph{linear} Dirac equation for $\psi$. Nevertheless, the following four elements 
make RDI much simpler than the traditional methods:
(i) The Dirac equation is written in the form where both $\psi$ and $A_{\mu}$ 
are $2 \times 2$ complex matrices \cite{BaylisClassicalSpinor1992,BaylisBook1996}.
(ii) The cross term responsible for the bilinearity is eliminated by expressing the vector 
potential as an explicit function of the state.          
(iii) The physical consistency of the state is accomplished by demanding the Hermiticity of 
the vector potential expressed in matrix form.
(iv) Enforcing the Lorentz covariance by decomposing the state into spacetime rotations as well as a transformation of the internal degrees of freedom significantly reduces the complexity of the analytic derivations.

\emph{Methodology of Relativistic Dynamical Inversion.}
The Dirac equation (\ref{Dirac-Equation}) can be written in different forms
emphasizing the geometry of the Lorentz group
\cite{hestenes1967real,hestenes1973local,hestenes1975observables,hestenes2003mysteries,hestenes2010zitterbewegung,lounesto2001clifford,doran2003geometric,BaylisClassicalSpinor1992,BaylisBook1996}.
Here, we employ the Baylis formulation \cite{BaylisClassicalSpinor1992,BaylisBook1996,BaylisClassicalSpinorEMW1999,baylis1999electrodynamicsBook,baylisQuantumClassical2010} 
(see also Sec. I of the Appendix) where the 
state $\psi$
in Eq. (\ref{Dirac-Equation}) is represented by the matrix $\Psi$ and its Clifford conjugate $\bar{ \Psi }$,  
\begin{align} \notag
\psi = \begin{pmatrix}
        \psi_1 \\ \psi_2 \\ \psi_3 \\ \psi_4
       \end{pmatrix}
\Longleftrightarrow
\left\{
{\Psi = \begin{pmatrix}
    \psi_1 + \psi_3  & - \psi_2^{*} + \psi_4^* \\
    \psi_2 +\psi_4    &  \psi_1^*       -\psi_3^*  
    \end{pmatrix},
 \atop
 \bar{ \Psi } = \begin{pmatrix}
   \psi_1^*  -\psi_3^*&   \psi_2^{*} - \psi_4^* \\
    -\psi_2 -  \psi_4    &   \psi_1 + \psi_3    
    \end{pmatrix}. }
    \right.
\end{align} 
obeying the Dirac equation in the matrix form \cite{BaylisClassicalSpinor1992,BaylisBook1996}
\begin{align}  \notag
   i c \hbar \bar{\partial} \Psi \sigma_3 - c e \bar{A} \Psi  -  m c^2 \bar{\Psi}^\dagger = 0, 
\end{align}
where  $\bar{A} = \sigma_{\mu}A_{\mu}$, $\bar{\partial} = \sigma_{\mu} \partial_{\mu} $,
$\sigma_0 = \boldsymbol{1}$ is an identity matrix, $\sigma_{1,2,3}$ are Pauli matrices. Note that $\bar{A}$ must be a Hermitian matrix by construction.
According to Ref. \cite{lounesto2001clifford},  $\det\Psi = 0$  for the Majorana and Weyl fermions as well as for the flag-dipole spinors, 
whereas $\det \Psi \neq 0$ for electrons/positrons. 
Thus, in the latter case, the vector potential may be expressed as a function of the state
\begin{align}
\label{A-inverted}
	c e{\bar{A}} = \left(ic\hbar \bar{\partial} \Psi \sigma_3 
- mc^2 \bar{\Psi}^{\dagger}  \right)\Psi^{-1}.
\end{align} 
A crucial insight is the spinor factorization for electrons/positrons: 
$\Psi =  \sqrt{\rho} L$, where $\rho$ is a non-negative scalar function modulating the probability density \footnote{
Stationary solutions of the Dirac equation rarely have nodes; 
as a result, they cannot be used to classify the eigensolutions.
For example, the hydrogen atom eigenstates have no zeros in $\rho$ (except at the origin) \cite{grant2007relativistic}; 
likewise, nodes in the Landau levels for the Dirac equation are hard to come across.
}
and $L$ is an invertible matrix representing a  
Lorentz group element \cite{hestenes1967real,hestenes1973local,hestenes1975observables}.

Considering that a member $L$ of the special Lorentz group \cite{hestenes1967real,hestenes1973local,
hestenes1975observables} 
is composed of spatial rotations $R$, a boost $B$ and a transformation
of internal degrees of freedom generated by the Yvon-Takabayashi 
angle $\beta$ \cite{yvon1940equations,takabayasi1957relativistic}, the 
state can be factorized as \cite{hestenes1967real,hestenes1973local,hestenes1975observables,
BaylisBook1996}
\begin{align}
\label{Psi-factorized}
 \Psi = \sqrt{\rho} \, B R  e^{i\beta/2}.  
\end{align}
The boost $B$ is parametrized by the velocity components $c\mathbf{u}=c(u^1,u^2,u^3)$ 
(bold symbols denote three dimensional vectors throughout)
\begin{align} \label{boost-u}
 B = B(\mathbf{u}) = \frac{  u^{\mu} \sigma_{\mu} +  \mathbf{1} }{ 
 \sqrt{2(1+u^0)} },
\end{align}
with $ u^0 =  \sqrt{1 + \mathbf{u}^2  } $; whereas, the spatial rotations are 
parametrized by the angles $   \boldsymbol{\theta} = ( \theta^1 , \theta^2, \theta^3 )$
\begin{align}
\label{R-rotation}
 R =  R( \boldsymbol{\theta} ) = \exp \left(  -i \theta^k \sigma_k/2  \right).
\end{align}
Note that the density $\rho$, velocity $\mathbf{u}$, 
rotation angle $\boldsymbol{\theta}$, and Yvon-Takabayashi angle $\beta$ are in general 
functions of space and time.

RDI is performed in the following way:
Spacetime functions $\rho$, $\mathbf{u}$, $\boldsymbol{\theta}$, and $\beta$
are initially selected to describe a desired dynamics of the Dirac state $\Psi$.
The constructed factorization (\ref{Psi-factorized}) is substituted in Eq. (\ref{A-inverted})
to obtain the vector potential in the matrix form $\bar{A}$.  

If $\bar{A}$ is not Hermitian,
the proposed dynamics is not reachable with physical fields, and the 
parametrization  $\rho$, $\mathbf{u}$, $\boldsymbol{\theta}$, and $\beta$ 
needs to be modified.

If $\bar{A}$ is Hermitian, then the procedure is completed: The obtained vector 
potential $A_{\mu} = {\rm Tr}\,(\bar{A}\sigma_{\mu})/2$ enables to recover the electromagnetic 
fields $F^{\mu \nu} = c\left( \partial^{\mu} A^{\nu} - \partial^{\nu} A^{\mu} \right)$ 
and the source $J^{\nu} = \partial_{\mu} F^{\mu \nu} / (\varepsilon_0 c)$ generating them.  
Provided the current $J^{\nu}$, the obtained fields $F^{\mu \nu}$ necessarily satisfy Maxwell's equations. 
Note that $J^{\nu}$ 
differs from the  current $J_D^{\mu} = {\rm Tr}\,(\Psi\Psi^{\dagger} \sigma_{\mu}) = \psi^{\dagger} \gamma^0 \gamma^{\mu} \psi$ 
emanating from the Dirac equation.

RDI is a trial-and-error procedure to find a suitable parametrization $\rho$, $\mathbf{u}$, $\boldsymbol{\theta}$, $\beta$ of 
the desired dynamics to yield a pair $A_{\mu}$,  $\Psi$ analytically satisfying the Dirac equation. In a 
general case, the obtained $A_{\mu}$ may have a complicated temporal and special profile hard to implement experimentally.

Furthermore, RDI has a very general foundation, which is applicable to interactions beyond electromagnetic, 
e.g., non-linear Dirac equations and scalar interactions coupling through 
the mass ($mc^2 \rightarrow mc^2 + V$) as shown below. The inversion procedures in Refs. \cite{kruger1993classical, Oertel-DiracInversion-2015} can be viewed as specialized cases of RDI.

\emph{Dispersionless rotation.} 
We now find an electromagnetic field that  
moves a Gaussian wave packet along a circular trajectory in the $x-y$ plane without distortion.
Since the center of the wave packet should follow the trajectory 
$\mathbf{r}(t) = r_0( \cos \omega t, \sin \omega t ,0) $,
the desired state evolution is
\begin{align} 
\label{circular-2D} 
 \Psi &=   e^{ -\frac{ e B_0 }{4  \hbar}[ (x-r_0 \cos \omega t)^2 + (y-r_0 \sin \omega t)^2 ] } 
   B( \mathbf{u} ), 
\end{align}
where $  \mathbf{u} = \dot{\mathbf{r}}/ \sqrt{1 - (\dot{\mathbf{r}}/c)^2 } $ and
the values of $r_0$ and $\omega$ must be selected such that $r_0 \omega < c$ to avoid superluminal propagation.
According to RDI, the vector potential generating the dynamics
consists of a constant homogeneous magnetic field $B_0$ perpendicular to a planar electric field 
with a spatial and temporal profile. 
However, for the frequency 
\begin{align}
\label{circular2D-omega}
\hbar \omega_0 = mc^2 - \sqrt{ (mc^2)^2 + 2 e B_0 c^2 \hbar  },
\end{align}
the electric field acquires a fixed spatial configuration rotating in time.
This expression for $\omega_0$ can be regarded as the cyclotron frequency corrected for quantum effects 
(see Sec. III of the Appendix).

In Fig.~\ref{fig:EMF1}, the crossed circles represent the homogeneous 
magnetic field perpendicular to the plane, and the electric field at the initial time 
$t=0$ is displayed as blue arrows in the $x-y$ plane. 
According to Sec. III  of the Appendix, these electromagnetic fields satisfy 
Maxwell's equations with an electric current but without free charges. The black diffused 
circle (centered at $x=2 \micrometer $ and $y=0$) depicts the initial Gaussian [Eq. (\ref{circular-2D})]
state whose shape is preserved 
during the rotation along the gray circular arrow. 
\begin{figure}
  \centering
      \includegraphics[width=0.7\hsize]{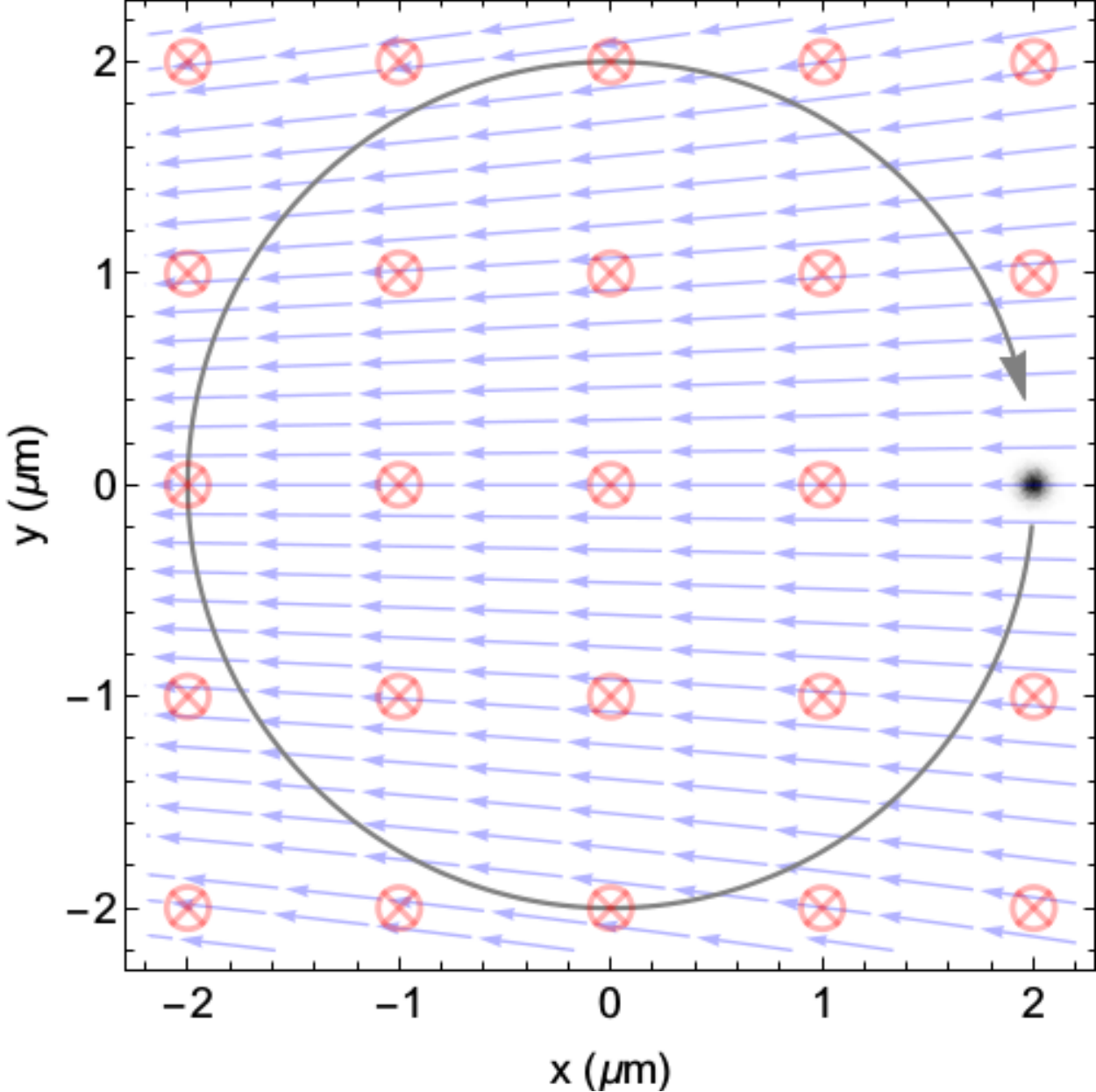}
      \caption{ Dispersionless rotation. The black diffused circle represents the electron cloud [Eq. (\ref{circular-2D})]  
      rotating along the circle with frequency $\omega$ without changing its shape.
      This dynamics is achieved by a combination of a rotating  electric field with a fixed spatial configuration
      (blue arrows) and a homogeneous magnetic field $B_0$ perpendicular to the plane  (crossed red circles). 
      The values of the parameters are $r_0=2 \micrometer $, $B_0=0.35$T, and $\omega =-61.55$ns$^{-1} $ obeying Eq. (\ref{circular2D-omega}).}
  \label{fig:EMF1}
\end{figure}

The nonrelativistic limit $c \to \infty $ of the driving controls consist of the homogeneous
magnetic field $B_0$ and the circularly polarized electric field:
$- (r_0 \omega/e) \left( \left(eB_0+m \omega \right) \cos \omega t, \left(eB_0+m \omega \right) \sin \omega t \right)$.
This setup can be shown to preserve the Gaussian shape within the Schr\"odinger equation. 

As shown in Appendix III \ref{Sec:Rotation2D}, the magnetic field is unaltered in the classical limit $\hbar \rightarrow 0$;
whereas, the vector norm difference between the exact electric field $\mathbf{E}$ and its classical limit  reads 
\begin{align}
 \left| \mathbf{E} -  \lim_{ \hbar \to 0 }{\mathbf{E}}  \right| = 
    \frac{\gamma r_0 \omega^3}{e}  \frac{\hbar}{2 c^2} 
\end{align}
where $\gamma = [1- (r_0 \omega/c)^2 ]^{-1/2}$ is the Lorentz factor. This reveals
that quantum effects are enhanced by relativistic dynamics.
The spatial inhomogeneity in the exact electric field depicted in Fig. \ref{fig:EMF1} 
is due to spin-orbit coupling, which is simultaneously a relativistic and quantum effect.
Note that this dynamics can be observed at experimentally 
available values of $B_0=0.35$T and $|\mathbf{E}| \sim 0.3$V/m employed in Fig.  \ref{fig:EMF1}. 
In such a regime, the synchrotron radiation energy loss per cycle is infinitesimally (i.e., $11$ orders of magnitude) 
smaller than the electron's kinetic energy. Therefore, the obtained solutions satisfy 
the physicality criterion. 

\begin{figure}
  \centering
      \includegraphics[width=1.\hsize]{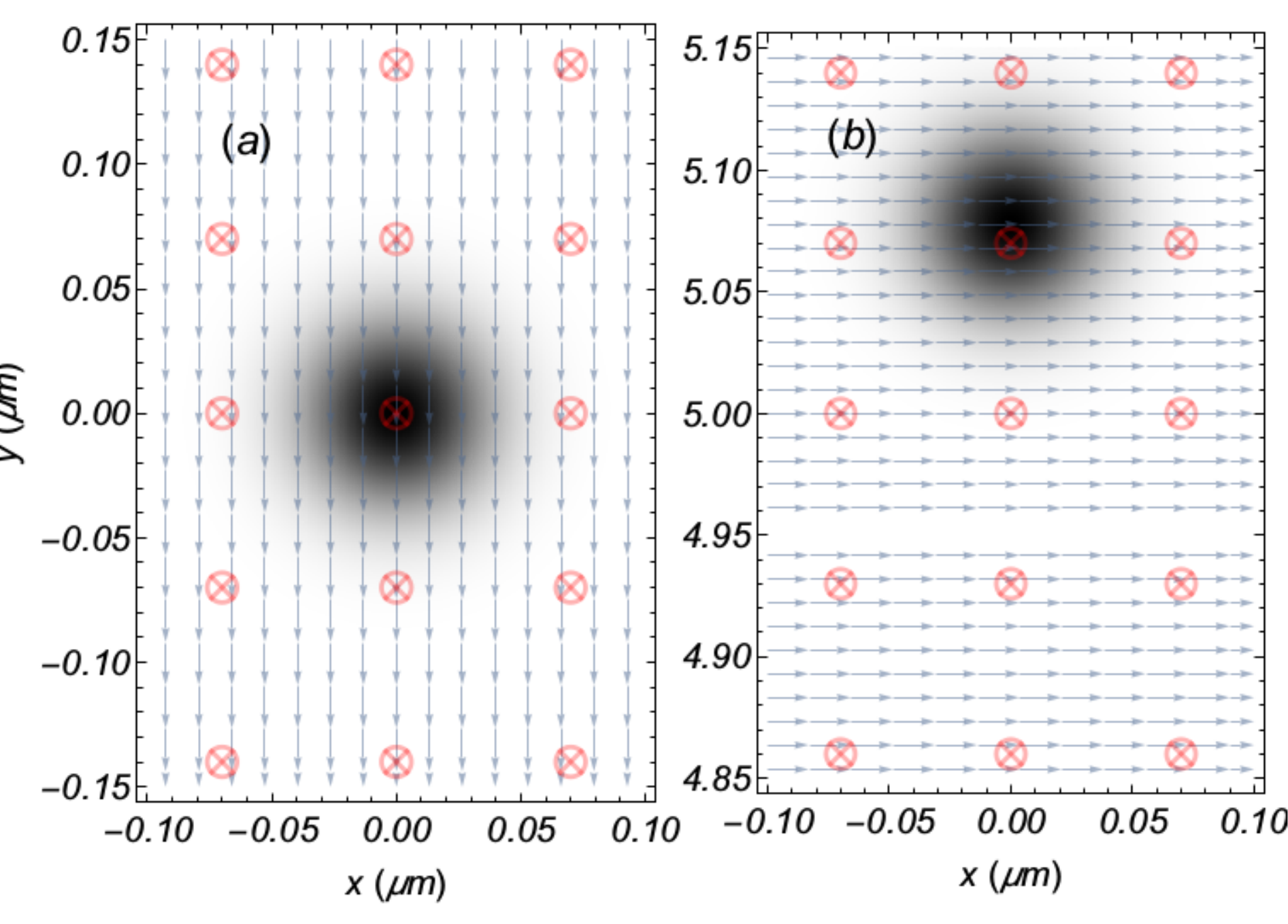}
      \caption{ Dispersionless translation of an electron. 
      Time snapshots of the state evolution [Eq. (\ref{translation-states})]  
      (a) at the beginning of the translation $t=0.$ps and (b) at $t=0.505$ns. The electromagnetic field in the Dirac equation 
performing this translation consists of the time-dependent homogeneous magnetic field 
perpendicular to the plane represented by red crossed circles while the electric field is displayed by blue arrows. 
The parameters in Eq. (\ref{translation-states}) are $L=10\, \mu$m, $T=1$ns and $B_0=1$T.  }
  \label{fig:translation-field}
\end{figure}

\emph{Dispersionless translation.}
We now apply RDI to achieve a spatial translation of a wave packet without changing its initial 
shape. For example, consider the translation along the $y$ axis with the trajectory $Y(t)$. 
Calculating the proper velocity $u$ from $\mathbf{r}(t)=(0,Y(t),0)$, 
we apply RDI to the dynamics
$ 
 \Psi =   e^{ -\frac{eB_0x^2}{4\hbar} } g(t,y) B( \mathbf{u} ) . 
$
It turns out that physical fields exist only if   
$ g(t,y) = G( y - Y(t)  )/ \sqrt{u^0(t)}$
for an arbitrary function $G(y)$. 
In particular, the translation of the Gaussian 
\begin{align}
\label{translation-states}
 \Psi &=   \frac{1}{\sqrt{u^0(t)}}\exp{\left( -\frac{eB_0 [x^2 + (y-Y(t))^2] }{4\hbar} \right)} 
 B( \mathbf{u} ) 
\end{align}
results in the electromagnetic field composed of a time dependent homogeneous magnetic 
field and an electric field with temporal and spatial dependence given in Sec. IV of the Appendix. 
For the specific trajectory $Y(t) = (L/2) [ 1 + \sin(\pi(t-T/2)/T  )]$
for $0 \leq t \leq T$,  Fig. \ref{fig:translation-field} 
displays two snapshots of the electric field at the beginning of motion 
[Fig. \ref{fig:translation-field}(a)] and 
at the middle [Fig. \ref{fig:translation-field}(b)]. 

In the nonrelativistic limit $c \to \infty$ the driving control is made of a constant 
magnetic field $B_0$ along $z$ and a time dependent electric field exclusively directed along the 
trajectory as dictated by Newton's law  $eE_2 =  m Y''(t)$. 

As elaborated in Sec. IV of the Appendix, the classical limit $\hbar \to 0$ affects
neither the magnetic field nor the electric field along the direction of motion. 
However, the exact component of the electric field perpendicular to the direction of motion can be written as
\begin{align}
eE_{1} = \left( \lim_{\hbar \to 0} eE_{1}  \right)  - \frac{\hbar}{ 2c^2 } \frac{d}{dt} \left( \gamma  \overset{..}{Y} \right),
\end{align}
where $\gamma = [1 - (\dot{Y}/c)^2]^{-1/2} $ is the Lorentz factor. 
This quantum correction resembles the Abraham-Lorentz force describing the interaction of 
a charged particle with its own electromagnetic field. Similar to the dispersionless rotation discussed above,
quantum effects are enhanced by the relativistic dynamics.
The counterintuitive temporal and spatial structure of the control shown in 
Fig. \ref{fig:translation-field}(b) is a manifestation of strong relativistic spin effects 
even at weak electric ($|\mathbf{E}| \sim 10^6 $ V/m) and magnetic ($B_0 \sim 1$T) fields. 
In this case, the bremsstrahlung energy loss is negligible compare to the electron's kinetic energy.


\emph{Integrable three dimensional solutions.}
Having demonstrated the RDI's ability to synthesize dynamics in two spatial dimensions, we now turn 
to  a challenging three dimensional case. For the following confined stationary state  
\begin{align} 
  \Psi = e^{   i  \arcsin[ f'(z) ]/2  } 
  e^{  -\frac{eB_0(x^2+y^2)}{4 \hbar}- mc f(z)/\hbar   }  e^{-i \epsilon t \sigma_3/ \hbar }, 
\label{Psi-3D}
\end{align}
RDI uncovers the underlying constant homogeneous magnetic field $B_0$ along the $z$ direction 
and the static electric potential 
\begin{align} \notag 
 eA_0 = \frac{ 2 mc (f'(z)^2-1) - \hbar f''(z)    }{2\sqrt{ 1 - f'(z)^2   }},
\end{align}
where the energy of the state  (\ref{Psi-3D})  is set to $\epsilon=0$, and $f(z)$ is an arbitrary 
real function. The obtained potential has no nonrelativistic limit. 

A noteworthy feature of the state (\ref{Psi-3D}) is the spatial dependence of the 
Yvon-Takabayashi angle $\beta = \arcsin f'(z) $, which is a signature of antiparticles represented 
by negative energy components in a wave packet (see, e.g., page 275 of Ref. 
\cite{BaylisBook1996}). 
The values of $\beta$ lie between $-\pi$ and $+\pi$, 
where particles (i.e., positive energy) and antiparticles 
are associated with $\beta=0$  and $\beta = \pm \pi$, respectively.
From the point of view of Lorentz transformations, 
the  Yvon-Takabayashi angle is a degree of freedom corresponding to the CPT conjugation \cite{lounesto2001clifford}
that includes the time inversion $t \to -t$; hence, $\beta$ 
is a parameter in the special Lorentz group not available in the restricted Lorentz group.  
Moreover, this degree of freedom is absent from the nonrelativistic Pauli-Schr\"odinger theory.
Since $f(z)$ controls the density of the state in Eq. (\ref{Psi-3D}), the tighter the confinement along 
the $z$ axis, the 
higher the contribution of antiparticles.

In the particular case of $ f(z) =   \sqrt{ \xi^2 + z^2 }$, where 
 $\xi$ determines the density spreading in $z$, the confining static electric potential is
the sum of soft-core Coulomb and short range potentials 
\begin{align} 
  e A_0 = - \frac{\xi m c}{ \sqrt{\xi^2 + z^2} } - \frac{\xi  \hbar}{ 2( \xi^2 + z^2  )  }. 
\label{soft-Coulomb}
\end{align}

In Sec. V of the Appendix, the space and time dependent electromagnetic fields are 
obtained by RDI to yield the dispersionless rotation of the state 
(\ref{Psi-3D}).

\emph{Exact solutions beyond electromagnetic interactions.}
RDI is not restricted to the electromagnetic interactions. 
The Dirac equation describing the scalar field $V$
coupled to the mass is 
$c \gamma^{\mu} \hat{p}_{\mu}  \psi = (mc^2 + V) \psi$.
This equation describes a Fermion in gravitational 
fields \cite{jentschura2014foldy}, topological materials \cite{shen2013topological},
and quark models \cite{chodos1974new,ru1984exact}. 
Another generalization of the Dirac equation involves nonlinear interactions \cite{thirring1958solubleAnnals,soler1970classicalPRD}, 
which can also be used to model Bose-Einstein condensates \cite{merkl2010chiralPRL}.

Let us consider the following nonlinear interaction with unspecified $V$
\begin{align}
\label{Dirac-nonlinear}
c \gamma^{\mu} \hat{p}_{\mu}  \psi = (mc^2 + V + \kappa |\psi|^2 ) \psi.
\end{align}
Applying RDI to the following state  
\begin{align} \label{Dirac-nonlinear-solution}
 \Psi = e^{i\pi/4} e^{ - m c z/ \hbar}e^{- mc z^2/(\xi \hbar)} e^{-i \epsilon t /\hbar \sigma_3}, 
\end{align}
we find the scalar interaction  $V = 2 m c^2 z /\xi - \kappa \sqrt{\frac{2 m c}{\pi \xi \hbar}} e^{-mc(2z+\xi)^2/(2\xi \hbar) }$ 
by demanding the absence of electromagnetic fields. 
Note that the state $\psi$ is confined in the potential $V$
unbounded  from above and below and, even more surprisingly, in the presence of an additional repulsive force emanating from the nonlinear term.
This is not possible in the nonrelativistic limit. Another example is presented
in Sec. VI of the Appendix.

These cases extend a rather short list
of analytic solutions of the Dirac equation with scalar interaction \cite{hiller2002solution,deCastro2003bound2D,deCastro2005exact}. 
Further explorations reveal that RDI becomes more flexible by utilizing both scalar and electromagnetic interactions, 
opening new possibilities for controlling quantum dynamics.

\emph{Outlook.}
We have developed RDI, a new framework for analytically constructing electromagnetic fields 
controlling the dynamics of the Dirac equation.  
RDI has also been shown to be a flexible tool for discovering novel exact solutions.
In particular, we have shown how relativistic coherent states could be constructed experimentally. 
A scalar interaction coupled to the mass has been incorporated into RDI.
This opens up prospects for quantum technologies
in new realms of physics and may further expand the scope of control landscape analysis \cite{rabitz2004quantum}.   

Since RDI relies on the matrix representation of the dynamical group 
generated by an equation of motion, the developed methodology may also be adaptable to other dynamical 
equations \cite{yepez2016quantum}. In a similar fashion, RDI may be used to yield exact solutions 
for non-Abelian fermions in the standard model \cite{trayling2001geometric} as well as curved 
spaces \cite{rodrigues2007many,fabbri2017torsion} that are currently intractable.

\emph{Acknowledgments.}
We thank two anonymous referees for a number of valuable suggestions. 
A.G.C. acknowledges financial support from NSF CHE 1464569, D.I.B. from DOE DE-FG02-02-ER-15344, R.
C. from ARO W911NF-16-1-0014 and H.R. from Templeton Foundation 52265. A.G.C. was also supported by the 
Fulbright Foundation.
D.I.B. is also supported by AFOSR Young Investigator Research Program (FA9550-16-1-0254).

A.G.C. and R.C. contributed equally to this work.

\onecolumngrid

\appendix

\section{I: Matrix form of the Dirac equation}
The traditional Dirac column spinor $\psi$ can be expressed in terms of $4\times 4$ complex matrices $\Psi$.
In particular, employing the Dirac matrix representation we have
\begin{align}
\psi = \begin{pmatrix}
        \psi_1 \\ \psi_2 \\ \psi_3 \\ \psi_4
       \end{pmatrix}
\Longleftrightarrow  \Psi=\begin{pmatrix}
    \psi_{1} & - \psi_2^{*} & \psi_3   & \psi_4^{*} \\
    \psi_2   &   \psi_1^{*} & \psi_4   & -\psi_3^* \\
    \psi_3   &   \psi_4^*  & \psi_1   & -\psi_2^* \\
    \psi_4   &  -\psi_3^*   & \psi_2   &  \psi_1^* 
    \end{pmatrix}, 
\end{align} 
where $ \Psi $ belongs to the group $Spin(1,3)$, as the double cover of the Special Lorentz group 
(at each point in the spacetime).
The spinor operator $\Psi$ obeys the Dirac-Hestenes 
equation \cite{hestenes1967real,hestenes1973local,hestenes1975observables,hestenes2003mysteries,hestenes2010zitterbewegung,lounesto2001clifford,doran2003geometric}
\begin{align}
   \hbar c \partial\!\!\!/ \Psi  \gamma^2\gamma^1 - c e A\!\!\!/ \Psi  -  m c^2 \Psi \gamma^0 = 0. 
   \label{Dirac-Hestenes-Eq}
\end{align} 
The Feynman slash notation is employed $A\!\!\!/ = A^{\mu}  \gamma_{\mu}$, $\partial\!\!\!/ = \gamma^{\mu} \partial_{\mu} $
($\mu,\nu = 0,1,2,3$), where the gamma matrices $\gamma^{\mu}$ are constructed to contain the Minkowski metric $g^{\mu \nu}$ 
according to
\begin{align}
   \frac{1}{2}(\gamma^{\mu}\gamma^{\nu} +\gamma^{\nu}\gamma^{\mu}) &= g^{\mu \nu} \boldsymbol{1}.
\end{align}
Other important properties of the gamma matrices are
\begin{align}
( {\gamma^0})^\dagger &= \gamma^0 ,\\
 (\gamma^k)^\dagger &= - (\gamma^k)^\dagger, \\
 \gamma^{\mu} &= \gamma_{\mu}^{-1}.
\end{align}
Among the infinite possibilities, the Dirac matrix representation is built in terms of the 
Kronecker product of Pauli matrices $\sigma_{\mu}$
\begin{align}
  \gamma^0   &= \sigma_3 \otimes \sigma_0 , \\
  \gamma^{j}  &= i \sigma_2 \otimes \sigma_j, 
\end{align}
where $\sigma_0 = \boldsymbol{1}$ and $j=1,2,3$. The Weyl representation is another possibility   
\begin{align}
 \gamma^0 &=    \sigma_1 \otimes \sigma_{0}, \\
 \gamma^{j}&=  i\sigma_2 \otimes \sigma_j.
\end{align}

\begin{figure}
  \centering
      \includegraphics[width=0.6\hsize]{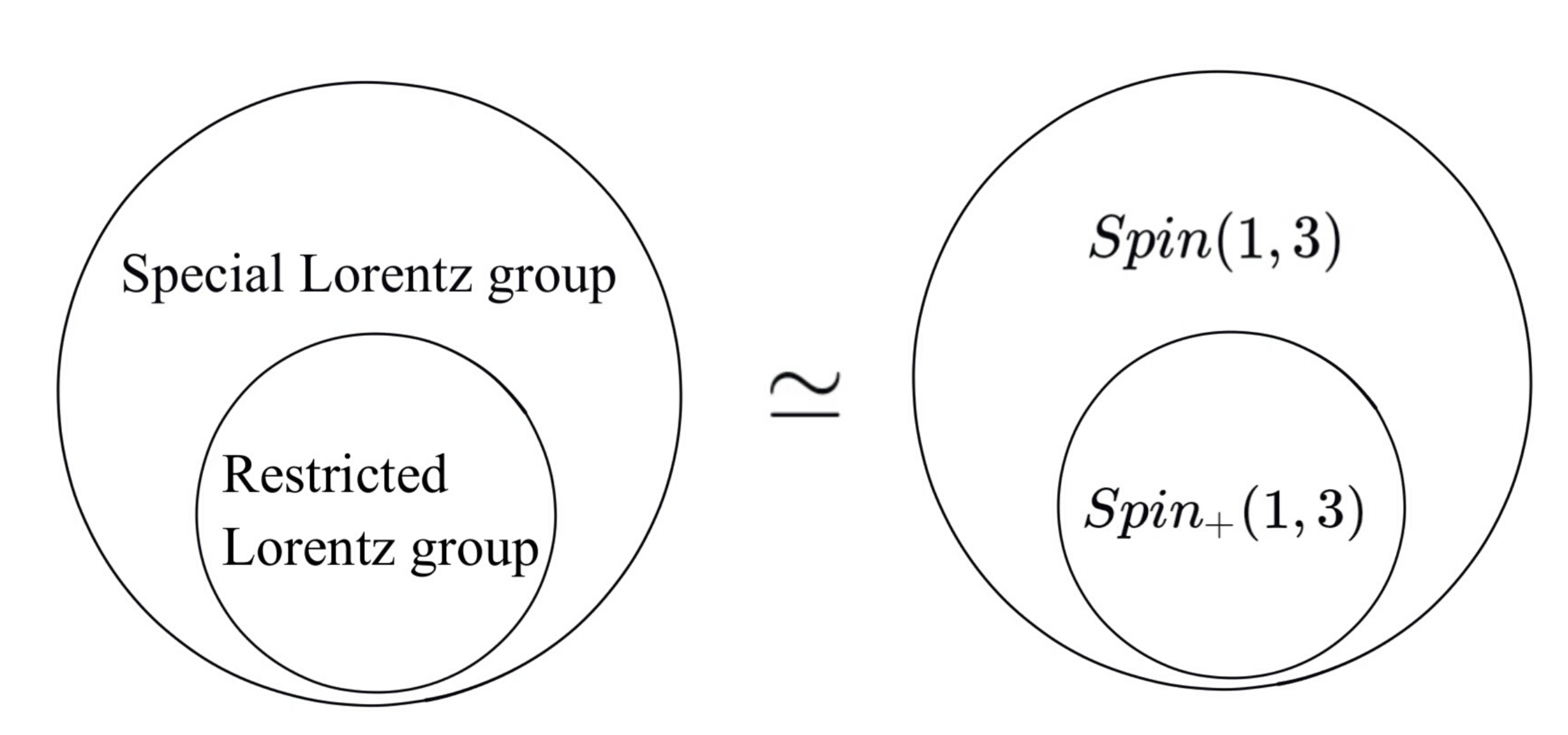}
      \caption{ Pictorial portrayal of the special Lorentz group and the restricted Lorentz group along 
      with their isomorphic representations in terms of the $Spin(1,3)$ and $Spin_{+}(1,3)$ groups.  }
\label{SpinAlgebraFig}
\end{figure}

The group  $Spin_{+}(1,3)$ is defined as the double cover of the \emph{restricted Lorentz group} 
and is characterized by preserving the direction of time while 
avoiding spatial reflections. A group element can be decomposed as
\begin{align}
 \Psi =  B R\,\, \in Spin_{+}(1,3), 
\end{align}
where $R$ and $B$ are unitary and Hermitian matrices, respectively. The matrix $R$ induces
spatial rotations while $B$ performs Lorentz boosts.

The double cover of the \emph{special Lorentz group} (allowing for $CPT$ conjugation) 
contains the additional factor $\exp \left( -i\gamma^5 \beta/2 \right)$ 
 \begin{align}
 \Psi =  \exp \left( -i\gamma^5 \beta/2 \right) R B \,\, \in Spin(1,3), 
\end{align}
where  $\beta$ is the  Yvon-Takabayashi angle \cite{yvon1940equations,takabayasi1957relativistic}.
The solutions of the Dirac equation (\ref{Dirac-Hestenes-Eq}) are spacetime modulations
of $Spin(1,3)$ carried out by a non-negative scalar function $\rho$ according to
\begin{align}
 \Psi =   \sqrt{\rho} \exp \left( -i\gamma^5 \beta/2 \right) R B. 
\end{align}

Alternative to the Dirac-Hestenes formulation, Dirac spinors can be 
expressed in terms of  $2\times 2$ complex matrices 
\begin{align}\label{EqBaylisDiracSpin}
\psi = \begin{pmatrix}
        \psi_1 \\ \psi_2 \\ \psi_3 \\ \psi_4
       \end{pmatrix}
\Longleftrightarrow
\Psi = \begin{pmatrix}
    \psi_1 + \psi_3  & - \psi_2^{*} + \psi_4^* \\
    \psi_2 +\psi_4    &  \psi_1^*       -\psi_3^*  
    \end{pmatrix},
\end{align} 
obeying the Dirac-Baylis equation \cite{BaylisClassicalSpinor1992,BaylisBook1996}
\begin{align}  
   i c \hbar \bar{\partial} \Psi \sigma_3 - c e \bar{A} \Psi  -  m c^2 \bar{\Psi}^\dagger = 0, 
\end{align} 
where $\sigma_3$ is a Pauli matrix that defines the arbitrary initial spin direction and 
the overbar is the Clifford conjugation that applied to a matrix $L$ gives
\begin{align}
  L = \begin{pmatrix}
       L_{11} & L_{12} \\ L_{21} & L_{22}
      \end{pmatrix}
 \to \bar{L} = 
      \begin{pmatrix}
       L_{22} & -L_{12} \\ -L_{21} & L_{11}
      \end{pmatrix}.
\end{align}
The gradient operator $\bar{\partial}$ and the vector potential $\bar{A}$ are  
\begin{align}
 \bar{\partial} =& \sigma_{\mu} \partial_{\mu} = \frac{1}{c} \sigma_0 \frac{\partial}{\partial t} + \sigma_1 \frac{\partial}{\partial x^1} +
  \sigma_2 \frac{\partial}{\partial x^2} + \sigma_3 \frac{\partial}{\partial x^3}, \\
  \bar{A} =& \sigma_0 A_0 + \sigma_1 A_1 + \sigma_2 A_2 + \sigma_3 A_3, 
\end{align}
where the physical components of the vector potential are given in contravariant indexes $A^{\mu}$ 
such that $A^0=A_0$ and $A^{k} = -A_{k}$.

Both Dirac-Hestenes and Dirac-Baylis approaches are completely equivalent.

A delocalized free particle with zero mean momentum and spin along $z$ can be represented as 
\begin{align}
\label{psi-rest}
\Psi_{rest} &=  \exp \left( - i  mc^2 t  \sigma_3 /\hbar \right),
\end{align}
where the spinor subscript indicates an specific reference frame that is taken at rest.

The spacetime position $x$ is 
\begin{align}
r &=  c t\sigma_0  + \mathbf{r}, \\
 \mathbf{r}  &= x \sigma_1  +  y \sigma_2  + z \sigma_3.  \nonumber
\end{align} 
The \emph{unitless} proper velocity $u$ is defined as
\begin{align}
 c u = \frac{d r }{d \tau},
\end{align}
where $\tau$ is the proper time (time in the rest frame attached to the particle).
Correspondingly, the proper velocity can be expressed as
\begin{align}
 u &=  \sigma_0 u^0 + \mathbf{u}, \\
 \mathbf{u}  &= \sigma_1 u^1 + \sigma_2 u^2 + \sigma_3 u^3 \nonumber,
\end{align} 
where the shell mass condition must be imposed by writing
$u^0 = \sqrt{  1 + ( u^1 )^2 + ( u^3 )^2 + ( u^3 )^2  } =  \sqrt{  1 + \mathbf{u}^2 }$,
effectively reducing the degrees of freedom of the proper velocity to three.
A state with a net velocity can be obtained through an active Lorentz transformation 
carried out by a Lorentz boost given by  
\begin{align}
  B( \mathbf{u} ) = \sqrt{ u },
\end{align} 
which can be written as
\begin{align}
  B( \mathbf{u}  ) = \frac{ u + \mathbf{1}  }{\sqrt{2(1+ u^0)}}.
\end{align}
For example, the spinor operator corresponding to a Lorentz boost along the $z$ direction 
can be expressed as 
\begin{align}
 B(0,0,u^3) =  \frac{1}{ \sqrt{ 2(1+u^0) } }
  \begin{pmatrix} 
     1 + u^0+u^3 & 0   \\
     0 & 1 + u^0 - u^3  
     \end{pmatrix},
\end{align}
with $u^0=\sqrt{ 1 + (u^3)^2 }$. 
The active Lorentz transformation of the coordinates is carried out by the following formula
\begin{align}
r \rightarrow r' = B(0,0,u^3) r B(0,0,u^3), 
\end{align}
leading to 
\begin{align} \label{boostedCoordinates}
  \begin{pmatrix}
   ct'^0 \\ z' 
  \end{pmatrix} =  
   \begin{pmatrix}
    u^0  & u^3 \\ u^3 & u^0  
  \end{pmatrix}
  \begin{pmatrix}
   c t^0 \\ z 
  \end{pmatrix}.
\end{align}

A state $\Psi(t,z)$ is actively boosted  to $\Psi'(t',z')$ according 
the spinorial transformation law that in the particular case reads
\begin{align}
 \Psi(t,z) \rightarrow 
 \Psi'(t',z') = B(0,0,u^3) \Psi(t,z),
\end{align}
where we must emphasize that the final transformed spinor $\Psi'$ must be expressed in terms of the boosted spacetime coordinates 
$t'$ and $z'$ according to Eq. (\ref{boostedCoordinates}).
Thus, the state in Eq. (\ref{psi-rest})  boosted along $z$ is 
\begin{align} 
 \Psi(t',z') =& B(0,0,u^3) \Psi_{rest}(t,z)\\
    =&  B(0,0,u^3) \exp 
\left[ -\frac{i m c}{\hbar} ( ct' u^0 - z' u^3 )\sigma_3  \right].
\end{align}
Considering that the Dirac equation is covariant under (homogeneous) restricted Lorentz transformations, 
$\Psi'$ must satisfy the Dirac equation.

\section{II: Time-dependent boost of a Gaussian state }
Let us initially consider the Landau ground state for a 
constant homogeneous magnetic field
 $  \mathbf{B}   = (0,0, B_0 )$
\begin{align}
 \label{rest-Gaussian}
  \Psi_{rest} = \exp \left( - \frac{B_0}{4 \hbar}( x^2 + y^2  )   \right)
  \exp \left( -\frac{i}{\hbar} m c^2 t \sigma_3   \right),
\end{align}
and a constant Lorentz boost along the $y$ direction with
\begin{align}
    \begin{pmatrix} c t \\ x \\ y \\ z \end{pmatrix} =
      \begin{pmatrix} 
        u^0 & 0 &-u^2 & 0 \\[6pt] 
            0 & 1 & 0  &  0                     \\[6pt] 
        -u^2  & 0 &  u^0  & 0  \\[6pt]
         0 & 0 & 0 & 1
      \end{pmatrix}
      \begin{pmatrix} 
       c t' \\ x' \\y' \\ z'
      \end{pmatrix}.
 \end{align}
The ground state is boosted to 
\begin{align}
\label{boosted-psi}
  \Psi_{rest} \rightarrow \Psi( t', x',y'  ) 
  =  \exp \left[  - \frac{B_0}{4 \hbar} \left( x'^2 + [ c t' u ^2 - u^0y'  ]^2  \right) \right]
    B(0,u^2,0)
    \exp \left[ -\frac{i m c}{\hbar} \left( ct u^0 - u^2 y' \right) \sigma_3   \right].
\end{align}
Applying RDI we obtain the fields 
\begin{align}
   \mathbf{B}_{boosted} &= \left( 0,0,u^0 B_0 \right), \\
   \mathbf{E}_{boosted} &= \left( -u^2 B_0  ,0,0 \right).
\end{align}
Independently, we can verify that the Lorentz transformations of the fields are
\begin{align}
 \mathbf{B}  \rightarrow \mathbf{B}' &= \left( 0,0,u^0 B_0 \right), \\
 \mathbf{E}  \rightarrow \mathbf{E}' &= \left( -u^2 B_0  ,0,0 \right).
\end{align}
Thus, demonstrating the covariance of the Dirac equation under (homogeneous) Lorentz transformations.

The Dirac equation is covariant for homogeneous Lorentz transformations that do not depend on time or position.
Nevertheless, a non-homogeneous Lorentz transformed spinor can still statisfy the Dirac equation but 
for a properly designed electromagnetic field. Let us illustrate this situation by analyzing the action 
of a time dependent Lorentz transform that corresponds to the action of a constant electric field $E_0$ 

\begin{align}
  B = \exp \left[ \frac{1}{2} \arcsinh \left( \frac{eE_0 t'}{mc}  \right)  \sigma_3  \right], 
\end{align}
That induces the following transformation of coordinates to the rest frame 
\begin{align}
    \begin{pmatrix}  t \\ x \\ y \\ z \end{pmatrix} =
      \begin{pmatrix} 
       t' \sqrt{1+ \left(\frac{E_0 t'}{mc}\right)^2 } - t' \frac{ E_0 y }{m c^2}  \\ x \\ 
       y' \sqrt{1+ \left(\frac{E_0 t'}{mc}\right)^2 } - \frac{E_0 t'^2}{m}  \\ z
      \end{pmatrix}.
\end{align}
Applying this Lorentz boost to the Gaussian spinor at rest in Eq. (\ref{rest-Gaussian}) we get 
\begin{align}
 \Psi' = B \Psi_{rest}.
\end{align}
Ignoring the primes in the variables, RDI provides 
\begin{align*}
eA_0 =& 
\frac{eB (eE) R^2 t x+2 c^4 m^4+4 c^2 (eE)^2 m^2 t^2-2 m \sqrt{Q} R}{2 c^2 m^2 R}-\frac{(eE) y \left(eB (eE) t x+2 c^2 m^2\right)}{2 c^3 m^2},
\\
eA_1 =& 
\frac{eB_0 (eE_0)^4 t^2 y^3}{2 c^6 m^4} - \frac{eB_0 (eE_0)^6 t^4 y^3}{2 c^5 m^4 \sqrt{Q}} - \frac{2 eB_0 (eE_0)^5 t^4 y^2}{c^5 m^4 R} + 
\frac{3 eB_0 (eE_0)^7 t^6 y^2}{2 c^4 m^4 \sqrt{Q} R} + \frac{5 eB_0 (eE_0)^4 t^4 y}{2 c^4 m^4} + 
  \\& \frac{eB_0 c^4 (eE_0) m^2 t^2}{2 \sqrt{Q} R} - \frac{3 eB_0 (eE_0)^6 t^6 y}{2 c^3 m^4 \sqrt{Q}} - \frac{eB_0 (eE_0)^5 t^6}{c^3 m^4 R}
 -\frac{eB_0 (eE_0)^4 t^2 y^3}{2 c^3 m^2 \sqrt{Q}} - \frac{3 eB_0 (eE_0)^3 t^2 y^2}{2 c^3 m^2 R}
 \\&  -\frac{eB_0 c^3 m^2 y}{2 \sqrt{Q}}+\frac{eB_0 (eE_0)^7 t^8}{2 c^2 m^4 \sqrt{Q} R}+\frac{5 eB_0 (eE_0)^5 t^4 y^2}{2 c^2 m^2 \sqrt{Q} R}+
 \frac{eB_0 c^2 (eE_0)^3 t^4}{\sqrt{Q} R}+\frac{eB_0 (eE_0)^2 t^2 y}{c^2 m^2}
 -\frac{2 eB_0 (eE_0)^4 t^4 y}{c m^2 \sqrt{Q}}- 
 \\& \frac{eB_0 (eE_0)^3 t^4}{c m^2 R}-\frac{eB_0 c (eE_0)^2 t^2 y}{\sqrt{Q}}+\frac{eB_0 (eE_0)^5 t^6}{m^2 \sqrt{Q} R} + 
 \frac{eB_0 (eE_0)^3 t^2 y^2}{\sqrt{Q} R}-\frac{(eE_0)^2 y \hbar }{2 c^4 m^2}+\frac{(eE_0)^4 t^2 y \hbar }{2 c^3 m^2 \sqrt{Q}}+
 \\& \frac{(eE_0)^3 t^2 \hbar }{c^3 m^2 R}-\frac{(eE_0)^5 t^4 \hbar }{2 c^2 m^2 \sqrt{Q} R}+\frac{(eE_0) \hbar }{2 c R}-\frac{(eE_0)^3 t^2 \hbar }{2 \sqrt{Q} R},
\\
eA_2 =&
\frac{eB_0 c^3 m^2 x}{2 \sqrt{Q}}-\frac{eB_0 (eE_0)^5 t^4 x y}{c^2 m^2 \sqrt{Q} R}+
\frac{eB_0 (eE_0)^4 t^2 x y^2 \sqrt{\frac{(eE_0)^2 t^2}{c^2 m^2}+1}}{2 c^2 m \sqrt{Q} R}+\frac{eB_0 (eE_0)^4 t^4 x}{2 c m^2 \sqrt{Q}}+
\\& \frac{eB_0 c (eE_0)^2 t^2 x}{2 \sqrt{Q}}-\frac{eB_0 (eE_0)^3 t^2 x y}{\sqrt{Q} R}+\frac{(eE_0)^2 t y}{c^2 m}-\frac{(eE_0)^3 t^3}{c m R}-\frac{c (eE_0) m t}{R}+(eE_0) t ,\\
eA_3 =& 
\frac{eB_0 (eE_0)^4 t^2 y^3}{2 c^6 m^4}-\frac{eB_0 (eE_0)^6 t^4 y^3}{2 c^5 m^4 \sqrt{Q}}-\frac{2 eB_0 (eE_0)^5 t^4 y^2}{c^5 m^4 R}+
\frac{3 eB_0 (eE_0)^7 t^6 y^2}{2 c^4 m^4 \sqrt{Q} R}+\frac{5 eB_0 (eE_0)^4 t^4 y}{2 c^4 m^4}+
\\& \frac{eB_0 c^4 (eE_0) m^2 t^2}{2 \sqrt{Q} R}-
 \frac{3 eB_0 (eE_0)^6 t^6 y}{2 c^3 m^4 \sqrt{Q}}-\frac{eB_0 (eE_0)^5 t^6}{c^3 m^4 R}-\frac{eB_0 (eE_0)^4 t^2 y^3}{2 c^3 m^2 \sqrt{Q}}
 -\frac{3 eB_0 (eE_0)^3 t^2 y^2}{2 c^3 m^2 R}
 \\& -\frac{eB_0 c^3 m^2 y}{2 \sqrt{Q}}+\frac{eB_0 (eE_0)^7 t^8}{2 c^2 m^4 \sqrt{Q} R}+\frac{5 eB_0 (eE_0)^5 t^4 y^2}{2 c^2 m^2 \sqrt{Q} R}
 +\frac{eB_0 c^2 (eE_0)^3 t^4}{\sqrt{Q} R}+\frac{eB_0 (eE_0)^2 t^2 y}{c^2 m^2}-\frac{2 eB_0 (eE_0)^4 t^4 y}{c m^2 \sqrt{Q}}
 \\& -\frac{eB_0 (eE_0)^3 t^4}{c m^2 R}-\frac{eB_0 c (eE_0)^2 t^2 y}{\sqrt{Q}}+\frac{eB_0 (eE_0)^5 t^6}{m^2 \sqrt{Q} R}+
 \frac{eB_0 (eE_0)^3 t^2 y^2}{\sqrt{Q} R}-\frac{(eE_0)^2 y \hbar }{2 c^4 m^2}+\frac{(eE_0)^4 t^2 y \hbar }{2 c^3 m^2 \sqrt{Q}}+
 \\& \frac{(eE_0)^3 t^2 \hbar }{c^3 m^2 R}-\frac{(eE_0)^5 t^4 \hbar }{2 c^2 m^2 \sqrt{Q} R}+\frac{(eE_0) \hbar }{2 c R}
 -\frac{(eE_0)^3 t^2 \hbar }{2 \sqrt{Q} R}.
\end{align*}
 with $ \text{Q} = c^6 m^4+c^4 {eE_0}^2 m^2 t^2+c^2 {eE_0}^4 t^4-2 c {eE_0}^3 t^2 y \sqrt{c^2 m^2+{eE_0}^2 t^2}+{eE_0}^4 t^2 y^2  $
 and $R=\sqrt{ (mc)^2  + (eEt)^2}$.
 The configuration of this vector potential is evidently not the Lorentz transform of the initial reference 
 vector potential. This becomes more obvious by observing that the complicated vector potential depends on the internal 
 degrees of freedom of the state and on $\hbar$.

\section{III: Dispersionless rotation: Gaussian state in 2D}\label{Sec:Rotation2D}
The state that describes a dispersionaless rotation of a Gaussian state is [Eq. (6) in the main text]
\begin{align} 
\label{circular-2D} 
 \Psi &=   \exp \left( -\frac{e B_0 }{4  \hbar}[ (x-r_0 \cos \omega t)^2 + (y-r_0 \sin \omega t)^2 ] \right) 
   B( \mathbf{u} ).
\end{align}
According to RDI, the electromagnetic vector potential is
\begin{align*}
 eA_0 =& 
 \frac{r_0 \omega  \left(eB_0+2 m \omega \right) (x \cos (t \omega )+y \sin (t \omega ))}{2 \sqrt{c^2-r_0^2 \omega ^2}}-\frac{r_0^2 \omega  \left(eB_0\right)}{2 \sqrt{c^2-r_0^2 \omega ^2}}+\frac{\omega  \hbar }{2 \sqrt{c^2-r_0^2 \omega ^2}}-\frac{\omega  \hbar }{2 c},
 \\
 eA_1 =& 
 \frac{y \left(eB_0\right) \left(-2 c^2+r_0^2 \omega ^2 \cos (2 t \omega )+r_0^2 \omega ^2\right)-2 r_0 \sin (t \omega ) \left(c^2 \left(-\left(eB_0\right)\right)+r_0 x \omega ^2 \left(eB_0\right) \cos (t \omega )+\omega ^2 \hbar \right)}{4 c \sqrt{\left(c-r_0 \omega \right) \left(c+r_0 \omega \right)}},
 \\
 eA_2 =& 
\frac{r_0 \left(\cos (t \omega ) \left(2 \omega ^2 \hbar -2 c^2 \left(eB_0\right)\right)+r_0 \omega ^2 \left(eB_0\right) (x \cos (2 t \omega )+y \sin (2 t \omega ))\right)+x \left(eB_0\right) \left(2 c^2-r_0^2 \omega ^2\right)}{4 c \sqrt{\left(c-r_0 \omega \right) \left(c+r_0 \omega \right)}},
 \\
 eA_3 =& 0.
\end{align*}

\begin{figure}
  \centering
      \includegraphics[width=0.4\hsize]{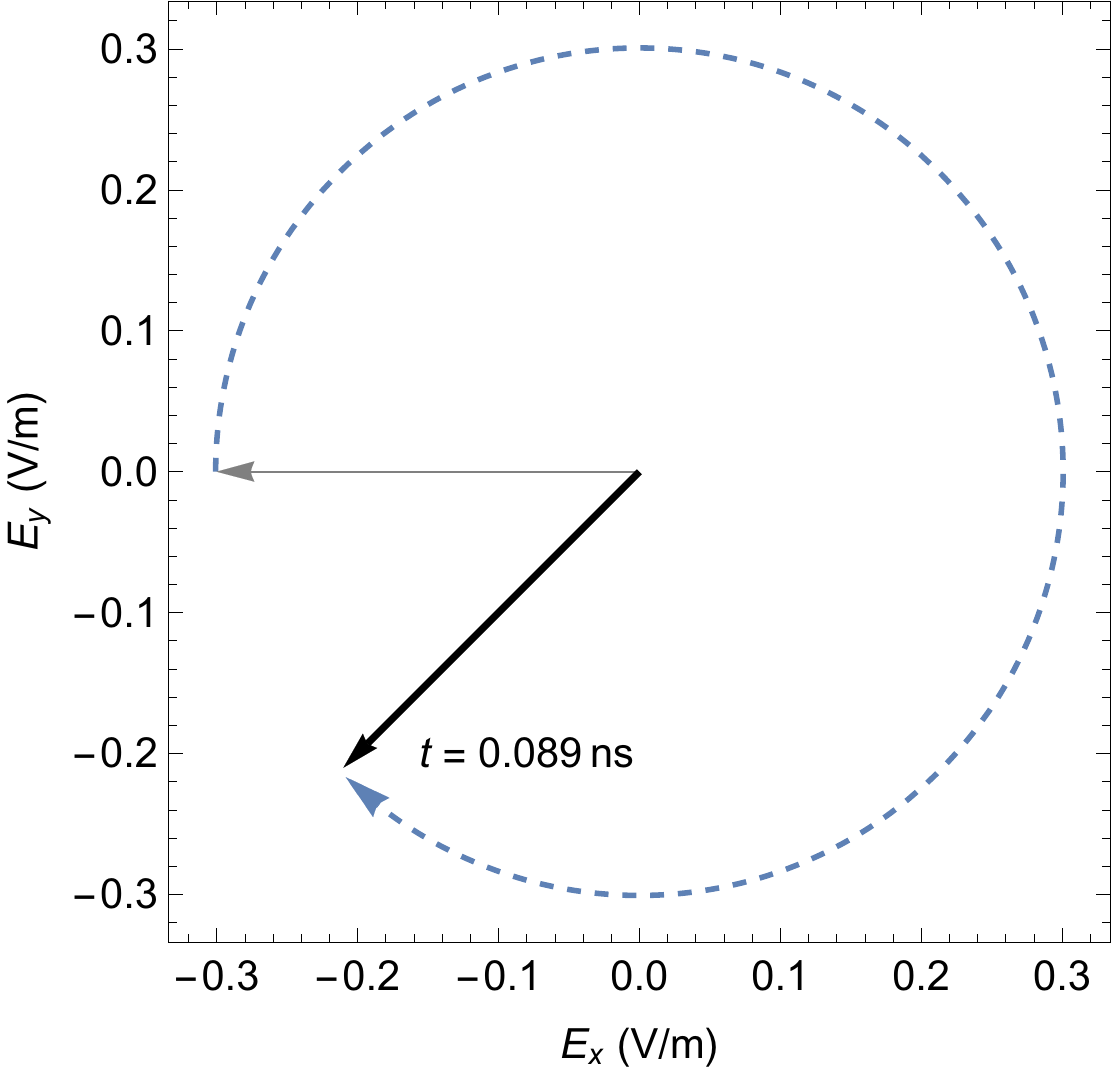}
      \caption{ Dispersionless rotation. 
The control electric field at $x=0,y=0, t= 0.089$ns (black arrow) as it rotates with the quantum corrected cyclotron frequency
      $\omega$ defined in Eq. (7) of the main text.  
      The values of the parameters are $r_0=2 \micrometer $, $B_0=0.35$T, and $\omega=-61.55$ns$^{-1} $.}
  \label{fig:EMF1-E}
\end{figure}

The corresponding electric field is (see Fig. \ref{fig:EMF1-E})
\begin{align}
 eE_1 =& 
  \frac{r_0 \omega  \left(\cos (t \omega ) \left(\omega ^2 \hbar -2 c^2 \left(eB_0+m \omega \right)\right)+r_0 \omega ^2 \left(eB_0\right) (x \cos (2 t \omega )+y \sin (2 t \omega ))\right)}{2 c \sqrt{(c-r_0 \omega ) (c+r_0 \omega )}},
 \label{RotationE1} \\
 eE_2 =& 
 \frac{r_0 \omega  \left(\sin (t \omega ) \left(-2 c^2 \left(eB_0+m \omega \right)+2 r_0 x \omega ^2 \left(eB_0\right) \cos (t \omega )+\omega ^2 \hbar \right)-r_0 y \omega ^2 \left(eB_0\right) \cos (2 t \omega )\right)}{2 c \sqrt{(c-r_0 \omega ) (c+r_0 \omega )}},
 \\
 eE_3 =& 0. \label{RotationE3}
\end{align}
The magnetic field is 
\begin{align}
 eB_1 =& 0, \label{RotationB1} \\
 eB_2 =& 0,\\
 eB_3 =&  \frac{\left(eB_0\right) \left(2 c^2-r_0^2 \omega ^2\right)}{2 c \sqrt{c^2-r_0^2 \omega ^2}}. \label{RotationB3}
 \end{align}

The obtained electromagnetic field obeys Maxwell's equations 
\begin{align}
 \nabla \cdot \mathbf{E} &= 0, \label{MaxwellEq1} \\
 \nabla \cdot \mathbf{B} &= 0, \\
 \nabla \times \mathbf{E} + \frac{\partial }{\partial t} \mathbf{B} &= 0, \\
 \nabla \times \mathbf{B} - \frac{1}{c^2} \frac{\partial}{\partial t} \mathbf{E} &=  \mu_0 \mathbf{J}, \label{MaxwellEq4}
\end{align}
with the current $\mathbf{J}$ 
\begin{align}
 \mu_0 e J_1 =& \label{RotationMaxwellJ1}
-\frac{r_0 \omega  \left(2 r_0 \omega ^3 \left(eB_0\right) (y \cos (2 t \omega )-x \sin (2 t \omega ))-\omega  \sin (t \omega ) \left(\omega ^2 \hbar -2 c^2 \left(eB_0+m \omega \right)\right)\right)}{2 c^3 \sqrt{c^2-r_0^2 \omega ^2}},
 \\
   \mu_0 e J_2 =&
-\frac{r_0 \omega ^2 \left(\cos (t \omega ) \left(\omega ^2 \hbar -2 c^2 \left(eB_0+m \omega \right)\right)+2 r_0 \omega ^2 \left(eB_0\right) (x \cos (2 t \omega )+y \sin (2 t \omega ))\right)}{2 c^3 \sqrt{c^2-r_0^2 \omega ^2}}, \\
\mu_0 e J_3 =& 0. \label{RotationMaxwellJ3}
\end{align}
The current amplitude
\begin{align}
 |\mu_0 e \mathbf{J}|^2 =& 
   \frac{r_0^2 \omega ^4}{4 c^6 (c^2-r_0^2 \omega^2 ) } \left(\sin (t \omega ) \left(\omega ^2 \hbar -2 c^2 \left(eB_0+m \omega \right)\right)-2 r_0 \omega ^2 \left(eB_0\right) (y \cos (2 t \omega )-x \sin (2 t \omega ))\right){}^2 
 \notag   \\ & + \frac{r_0^2 \omega ^4}{4 c^6 (c^2-r_0^2 \omega^2 ) }  
    \left(\cos (t \omega ) \left(\omega ^2 \hbar -2 c^2 \left(eB_0+m \omega \right)\right)+2 r_0 \omega ^2 \left(eB_0\right) (x \cos (2 t \omega )+y \sin (2 t \omega ))\right){}^2.
 \end{align}
 becomes time independent if the condition $2 c (eB_0+c m \omega )-\omega ^2 \hbar =0$ is employed.
The frequency accomplishing this is
\begin{align} 
\label{frequency-circle2D}
 \omega_0 = \frac{1}{\hbar} ( mc^2 - \sqrt{(mc^2)^2  + 2 c^2 e B_0 \hbar }  ).
\end{align}
The following series expansion indicates that this contains quantum corrections to the 
proper cyclotron frequency $\omega_{c} = eB_0/m$
\begin{align}
\label{frequency-expansion-circle2D}
 \omega_0 = -\omega_c + \frac{\omega_c^2}{2mc^2} \hbar - \frac{\omega_c^3}{2(mc^2)^2} \hbar^2 + O(\hbar^3).
\end{align}
In classical mechanics, the cyclotron frequency $\omega$ and the proper cyclotron frequency $\omega_c$
are related by $\omega = \omega_c/\gamma $, thus
\begin{align}
 \omega = - \frac{c \,\omega_c}{ \sqrt{c^2 + (r_0 \omega_c)^2} } = -\omega_c + 
 \frac{r_0^2 \omega_c^3}{ 2 } \frac{1}{c^2} - \frac{3 r_0^4 \omega_c^5}{8} \frac{1}{c^4} + O\left( \frac{1}{c^6} \right),
\end{align}

The low energy expansion of the current at the resonant frequency (\ref{frequency-circle2D}) is  
\begin{align}
 \mu_0 e J_1 =&
  -\frac{r_0^2 \left(eB_0\right){}^5 \left(x \sin \left(\frac{2 t \left(eB_0\right)}{m}\right)+y \cos \left(\frac{2 t \left(eB_0\right)}{m}\right)\right)}{c^4 m^4} + O\left( \frac{1}{c^6} \right), \\
 \mu_0 e J_2 =&
-\frac{r_0^2 \left(eB_0\right){}^5 \left(x \cos \left(\frac{2 t \left(eB_0\right)}{m}\right)-y \sin \left(\frac{2 t \left(eB_0\right)}{m}\right)\right)}{c^4 m^4}  + O\left( \frac{1}{c^6} \right),\\
  \mu_0 e J_2 =& 0.
\end{align}

The non-relativistic limit of the electromagnetic field (\ref{RotationE1})-(\ref{RotationB3}) is 
\begin{align}
 e\mathbf{E}_{nr} &= 
  \left\{-r_0 \omega  \cos (t \omega ) \left(eB_0+m \omega \right),-r_0 \omega  \sin (t \omega ) \left(eB_0+m \omega \right), 0\right\}, \\
 \mathbf{B}_{nr} &= \{ 0,0,B_0 \}.
\end{align}

The classical limit $\hbar \rightarrow 0 $ of the electric field (\ref{RotationE1})-(\ref{RotationE3}) reads
\begin{align}
 eE_{1\, (classical)} &=  
 \frac{r_0 \omega  \left(r_0 \omega ^2 \left(eB_0\right) (x \cos (2 t \omega )+y \sin (2 t \omega ))-2 c^2 \cos (t \omega ) \left(eB_0+m \omega \right)\right)}{2 c \sqrt{c^2-r_0^2 \omega ^2}},
\label{RotationEclass1} \\
 eE_{2\, (classical)} &= 
 -\frac{r_0 \omega  \left(2 \sin (t \omega ) \left(c^2 \left(eB_0+m \omega \right)-r_0 x \omega ^2 \left(eB_0\right) \cos (t \omega )\right)+r_0 y \omega ^2 \left(eB_0\right) \cos (2 t \omega )\right)}{2 c \sqrt{c^2-r_0^2 \omega ^2}},
 \\
 eE_{3\, (classical)} &= 0. \label{RotationEclass3}
\end{align}

The difference between the exact (\ref{RotationE1})-(\ref{RotationB3}) and its classical limit (\ref{RotationEclass1})-(\ref{RotationEclass3}) is 
\begin{align}
 e|\mathbf{E} - \mathbf{E}_{classical}| = \frac{r_0 \omega^3 \hbar}{  2\sqrt{c^4-(c r_0 \omega)^2} } =
     \frac{  \gamma r_0 \omega^3 \hbar}{2 c^2}, \quad \gamma = \frac{1}{\sqrt{1- (r_0 \omega/c)^2 }}.
\end{align}
The leading term in $\hbar$ can be expressed in terms of the resonant frequency (\ref{frequency-circle2D})
\begin{align}
 e|\mathbf{E} - \mathbf{E}_{classical}| =& -\frac{(eB_0)^3 r_0 \hbar}{2c^2m^3} + O(\hbar^2)  \notag \\
    =& - e B_0 \omega_1 r_0 + O(\hbar^2), \label{omega1-Eq}
\end{align}
where $\omega_1 = \frac{\omega_c^2}{2mc^2} \hbar $ is the first quantum correction 
to the cyclotron frequency in Eq. (\ref{frequency-expansion-circle2D}).
Note that Eq. (\ref{omega1-Eq}) can be interpreted as a Lorentz force $ e \mathbf{v} \times \mathbf{B}.$

The current constructed from the Dirac spinors is $J_D = \Psi \Psi^{\dagger}$, whose components $J_D^{\mu} = Tr(\Psi\Psi^{\dagger} \sigma_{\mu})$ are
\begin{align}
 J_D^0 =&\frac{c \exp \left(-\frac{\left(eB_0\right) \left(\left(x-r_0 \cos (t \omega )\right){}^2+\left(y-r_0 \sin (t \omega )\right){}^2\right)}{2 \hbar }\right)}{\sqrt{c^2-r_0^2 \omega ^2}}, \label{EqJD1} \\
J_D^1 =& -\frac{r_0 \omega  \sin (t \omega ) \exp \left(-\frac{\left(eB_0\right) \left(\left(x-r_0 \cos (t \omega )\right){}^2+\left(y-r_0 \sin (t \omega )\right){}^2\right)}{2 \hbar }\right)}{\sqrt{c^2-r_0^2 \omega ^2}},\\
J_D^2 =& \frac{r_0 \omega  \cos (t \omega ) \exp \left(-\frac{\left(eB_0\right) \left(\left(x-r_0 \cos (t \omega )\right){}^2+\left(y-r_0 \sin (t \omega )\right){}^2\right)}{2 \hbar }\right)}{\sqrt{c^2-r_0^2 \omega ^2}},\\
 J_D^2 =& 0. \label{EqJD4}
\end{align}

The velocity associated with the current $J_D$ is obatined as $v^{k} = c J_D^{k}/ J_D^0$
\begin{align}
  v^1 =& -r_0 \omega \sin(\omega t), \\
  v^2 =&  r_0 \omega \cos(\omega t),
\end{align}
with the magnitude given by $|v| = r_0 \omega$. 
Thus, superluminal propagation is avoided if $r_0\omega <c$. Furthermore, the latter inequality is also obtained from the RDI consistency condition for the electromagnetic fields as described in the main text.

The current $\mathbf{J}$ (\ref{RotationMaxwellJ1})-(\ref{RotationMaxwellJ3}) entering Maxwell's equations (\ref{MaxwellEq1})-(\ref{MaxwellEq4}) is related to the current $\mathbf{J}_D$ (\ref{EqJD1})-(\ref{EqJD4})  constructed from the Dirac spinor in the following way: The Maxwell current $\mathbf{J}$ creates the electromagnetic field (\ref{RotationE1})-(\ref{RotationB3}) steering the Dirac wave packet (\ref{circular-2D}). Moving along a circular trajectory, a Dirac electron yields the current $\mathbf{J}_D$ emitting synchrotron radiation. If the radiation losses are large, the proposed control protocol may not work. Therefore, the control electromagnetic fields are physically meaningful if the electron kinetic energy is much larger than the energy emitted via synchrotron radiation. The dispersionless rotation shown in Fig. 1 of the main text obeys well this criterion because the radiative energy loss per period is $\propto 10^{-32}$J, whereas the electron kinetic energy is $\propto 10^{-21}$J.

\section{IV: Dispersionless translation }

The state undergoing dispersionless translation is [Eq. (9) of the main text]
\begin{align}
\label{translation-states}
 \Psi &=   \frac{1}{\sqrt{u^0(t)}}\exp{\left( -\frac{eB_0 [x^2 + (y-Y(t))^2] }{4\hbar} \right)} 
 B( \mathbf{u} ),
\end{align}
where $c \mathbf{u} = {\{} 0, c u^2  ,0 {\}} = {\{}0,\dot{Y}/\sqrt{ 1 - (\dot{Y}/c)^2  } ,0 {\}}  $ and $u^0(t) = 1/\sqrt{ 1 - (\dot{Y}/c)^2  }$.
RDI leads to the following vector potential components
\begin{align*}
eA_0 =&  \frac{c \left(4 c \left(\sqrt{c^2-\dot{Y}^2}+c\right)-\dot{Y}^2 \left(\sqrt{1-\frac{\dot{Y}^2}{c^2}}+3\right)\right) \left(x \dot{Y} \left(c^2-\dot{Y}^2\right) \left(eB_0\right)-2 c^2 m \overset{..}{Y} \left(y-t \dot{Y}\right)\right)}{2 \left(c \left(\sqrt{c^2-\dot{Y}^2}+c\right)-\dot{Y}^2\right)^3},
\\
eA_1 =&  \frac{\hbar  \overset{..}{Y}-\left(c^2-\dot{Y}^2\right) \left(eB_0\right) (y-Y(t))}{2 c \sqrt{c^2-\dot{Y}^2}},
\\
eA_2 =&  \frac{c x \left(eB_0\right)}{2 \sqrt{c^2-\dot{Y}^2}},
\\
eA_3 =&  0.
\end{align*}
The corresponding electromagnetic fields are
\begin{align}
  \mathbf{B} &=   \left\{0,0,\frac{B_0 \left(2 c^2-\dot{Y}^2\right)}{2 c \sqrt{c^2-\dot{Y}^2}}\right\} = 
   \left\{ 0,0, \frac{\gamma +  \gamma^{-1} }{2} B_0  \right\}
  ,  \label{TranslationEB1}
  \\
  eE_1 &= 
  \frac{\dot{Y}^3 eB_0 \left((y-Y) \overset{..}{Y}+3 c^2\right)-\dot{Y} \left(c^2 (y-Y) \overset{..}{Y} eB_0+\hbar  \overset{..}{Y}^2+2 c^4 eB_0\right)-c^2 \hbar  \overset{...}{Y}+\dot{Y}^2 \hbar  \overset{...}{Y} - \dot{Y}^5 eB_0}{2 c \left(c^2-\dot{Y}^2\right)^{3/2}},
  \\
  eE_2 &= 
  \frac{\overset{..}{Y} \left(2 c^3 m-c x \dot{Y} \left(eB_0\right)\right)}{2 \left(c^2-\dot{Y}^2\right)^{3/2}} ,
  \\
  eE_3 &= 0, \label{TranslationEB4}
\end{align}
where $\gamma$ is the Lorentz factor.

The nonrelativistic limit $c\to \infty$ is 
\begin{align}
  \mathbf{B}_{(nr)} &= \{0,0,B_0\},  \\ 
  e\mathbf{E}_{\, (nr)} &=  \frac{ \pi L }{2T}  \left\{-  eB_0 \sin \left(\frac{\pi  t}{T}\right) , 
    (\pi m/T) \cos \left(\frac{\pi  t}{T} \right)  ,0\right\} . 
\end{align}

The classical limit $\hbar \to 0$ is
\begin{align}
 eE_{1\, (classical)}  =&  
 \frac{\dot{Y} \left(eB_0\right) \left((Y-y) \overset{..}{Y}-2 c^2+\dot{Y}^2\right)}{2 c \sqrt{c^2-\dot{Y}^2}}, \\
 eE_{2\, (classical)}  =& \frac{c \overset{..}{Y} \left(2 c^2 m-B_0 x \dot{Y}\right)}{2 \left(c^2-\dot{Y}^2\right)^{3/2}},\\
  eE_{2\, (classical)}  =& 0.
\end{align}
This means that only the perpendicular component of the electric field includes quantum corrections. Moreover, it can be shown that 
\begin{align}
 eE_{1\, (classical)} - eE_{1} = \frac{d}{dt} \left(  \frac{\hbar  \overset{..}{Y}}{2 c \sqrt{c^2-\dot{Y}^2}}  \right) 
  =  \frac{\hbar}{ 2c^2 } \frac{d}{dt} \left( \gamma  \overset{..}{Y} \right).
\end{align}

\begin{figure}
  \centering
      \includegraphics[width=0.4\hsize]{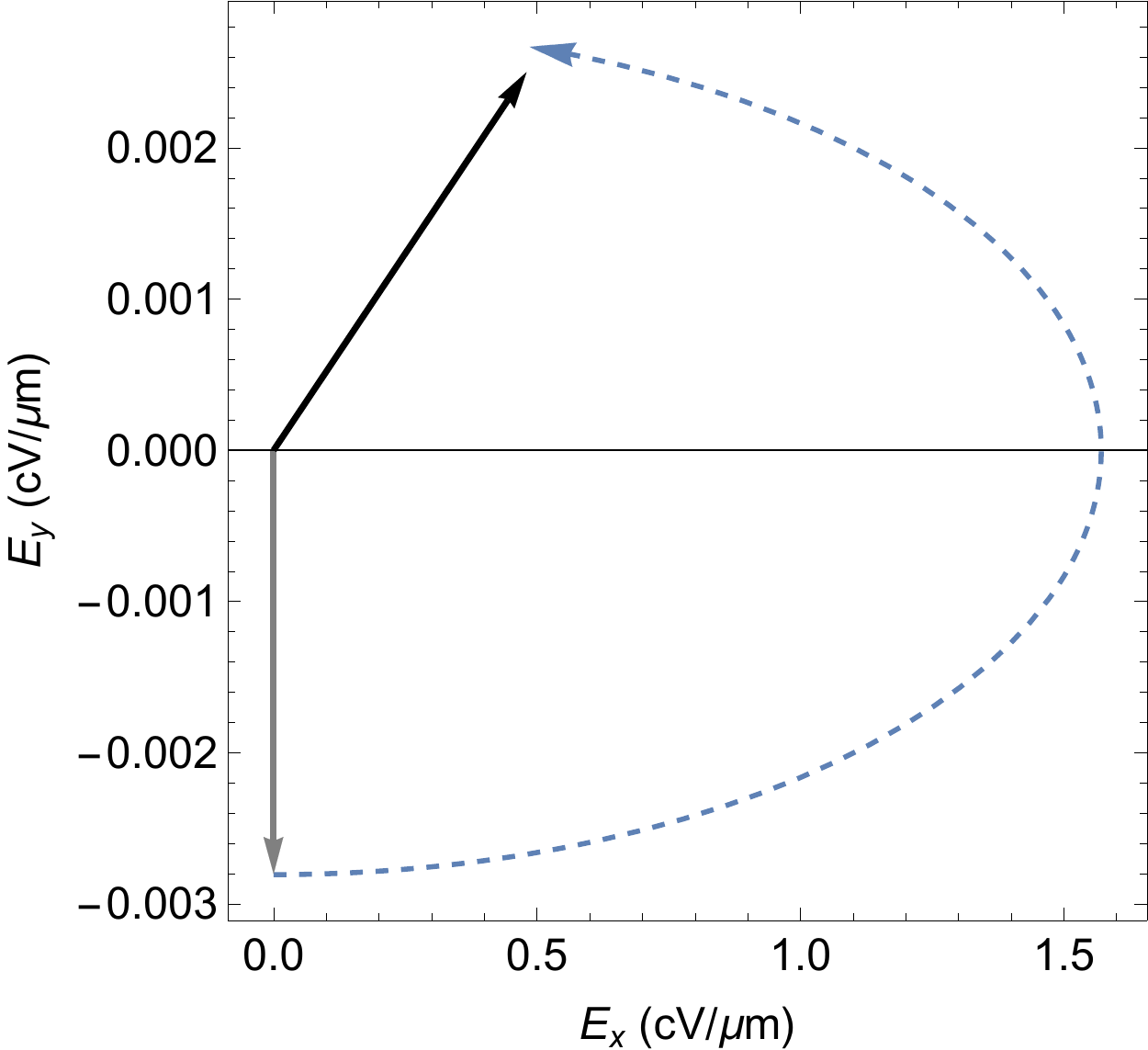}
      \caption{ Electric field at the origing as a function of time corresponding to Fig. 2 in the main text. 
      The parameter values  are found in Fig. 2 of the main text for the trajectory $Y(t) = (L/2) ( 1 + \sin(\pi(t-T/2)/T  )$.    }
  \label{fig:translationE}
\end{figure}

The electric field for the trajectory employed in the main text is
evaluated at the origin as a function of time and shown in Fig. \ref{fig:translationE}.

\begin{figure}
  \centering
      \includegraphics[width=0.4\hsize]{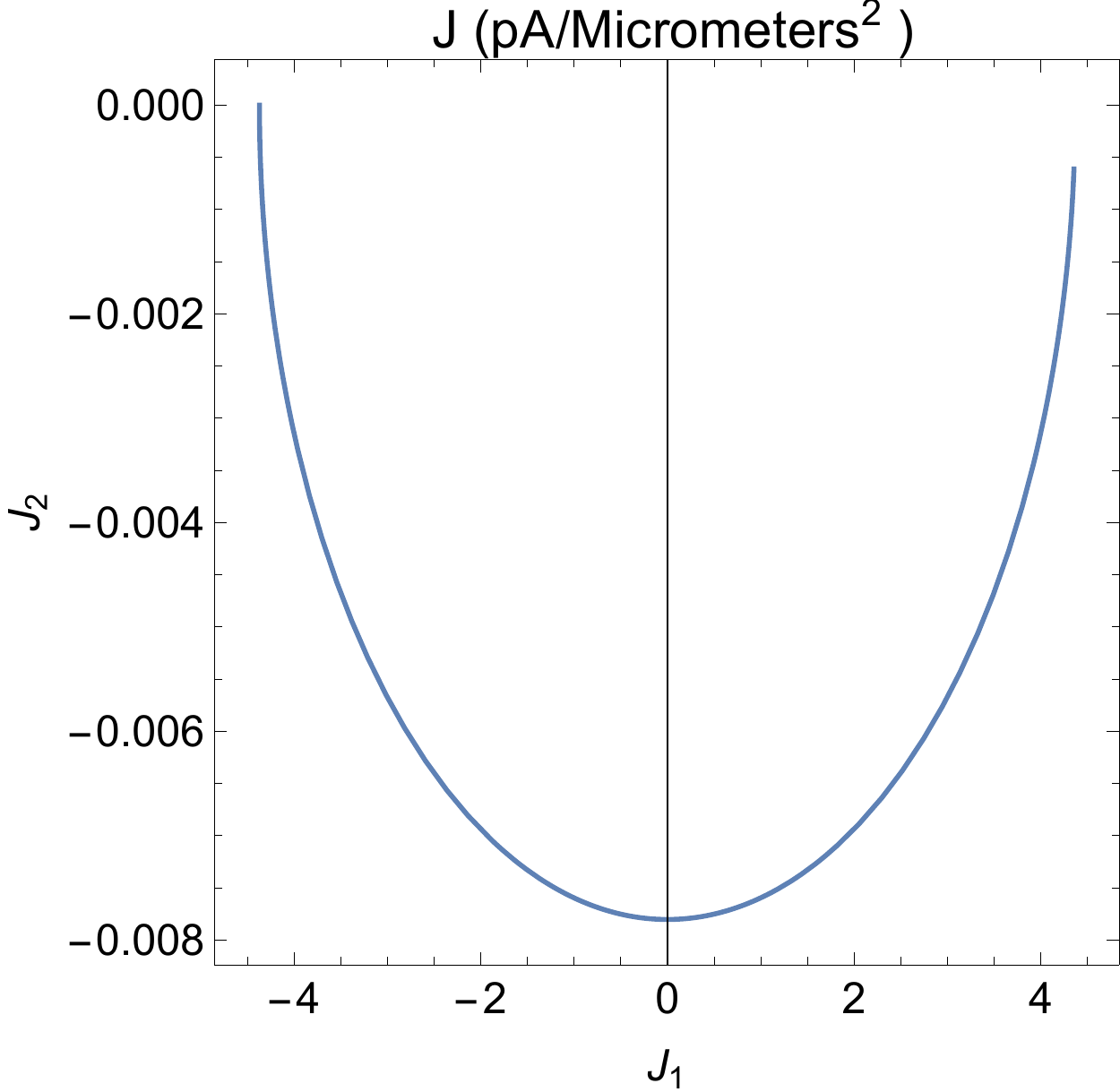}
      \caption{ Parametric plot of the current field as a function of time 
                in Eq. (\ref{Eq:TranslationJ}) required to generate the electromagnetic 
                field for the dispersionless translation of a wave packet. 
                The current at time $t=0$ is at the top left corner, 
                 while the current at time $t=9.75$ns is at the top right corner. 
                 The values of the parameters are found in Fig. 2 of the main text.   }
  \label{fig:TranslationJ}
\end{figure}

The electromagnetic fields (\ref{TranslationEB1})-(\ref{TranslationEB4}) obey Maxwell's equations (\ref{MaxwellEq1})-(\ref{MaxwellEq4}) with the following electric current:
\begin{align}
\mu_0 e J_1 &= \frac{Q^2 \dot{Y} \overset{...}{Y} \left(3 \hbar  \overset{..}{Y}+Q^2 (y-Y)
   \left(eB_0\right)\right)+3 Q^2 \dot{Y}^4 \overset{..}{Y} \left(eB_0\right)+2
   \dot{Y}^2 \hbar  \overset{..}{Y}^3+Q^4 \hbar  Y^{(4)}(t)}{2 c^3 Q^5}+
   \frac{c  \overset{..}{Y} \left(eB_0\right)}{Q^3} \label{TranslationJ1} \\
   & +\frac{\overset{..}{Y} \left(Q^2 (y-Y)
   \overset{..}{Y} \left(eB_0\right)+\hbar  \overset{..}{Y}^2-4 Q^2 \dot{Y}^2
   \left(eB_0\right)\right)}{2 c Q^5}, \\
 \mu_0 e J_2 &= \frac{c \left(x \overset{..}{Y}^2 \left(eB_0\right)-2 m Q^2 \overset{...}{Y}-6 m
   \dot{Y} \overset{..}{Y}^2\right)}{2 Q^5}+\frac{x \dot{Y} \left(eB_0\right)
   \left(Q^2 \overset{...}{Y}+2 \dot{Y} \overset{..}{Y}^2\right)}{2 c Q^5}, \\
 \mu_0 e J_2 &= 0, \label{TranslationJ3}
\end{align}
where $Q=\sqrt{c^2- (\dot{Y})^2 }$. The leading terms of the electric current at low energies are homogenous 
in the space 
\begin{align}
\label{Eq:TranslationJ}
 \mu_0 e \mathbf{J} = \left\{ \frac{1}{c^2}eB_0 \overset{..}{Y} + O\left( \frac{1}{c^4} \right) 
  , -\frac{1}{c^2} m \overset{...}{Y} +  O\left( \frac{1}{c^4} \right), 0 \right\}.
\end{align}
Figure \ref{fig:TranslationJ} depicts parametrically the current (\ref{TranslationJ1})-(\ref{TranslationJ3}) creating the fields (\ref{TranslationEB1})-(\ref{TranslationEB4}) that realize the dispersionless translation shown in Fig. 2 of the main text.

An electron moving linearly with an acceleration emits bremsstrahlung.  Similarly to the rotational dynamics discussed in Sec. \ref{Sec:Rotation2D}, the control electromagnetic fields (\ref{TranslationEB1})-(\ref{TranslationEB4}) are physically meaningful if the electron kinetic energy is much larger than the bremsstrahlung energy loss. The dispersionless translation depicted in Fig. 2 of the main text satisfies well this criterion since  the radiative energy loss does not exceed $10^{-38}$J while the electron kinetic energy  $\propto 10^{-24}$J.

\section{V: An integrable three-dimensional solution}
 The rotation of the three-dimensional state in Eq. (11) of the main text state is carried out 
 by 
\begin{align}
\Psi = &  e^{  -\frac{B_0}{4  \hbar}[(x-r_0 \cos \omega t)^2 + (y-r_0 \sin \omega t)^2]  }   
  e^{ - i \arcsin[ f'(z) ]/2 }  e^{  - mc f(z)/\hbar }    B( \mathbf{u} ) ,
\label{Eq:Psi-circular-3D}
\end{align}
where $B( \mathbf{u} ) $ is the same boost as in Eq. (\ref{circular-2D}). 

The vector potential according to RDI is
\begin{align*}
e A_0 =&   \frac{\sqrt{c^2-r_0^2 \omega ^2}}{2 c W^4 \sqrt{1-f'(z)^2}} \bigg{(}
-e B_0 c r_0 \omega  \sqrt{1-f'(z)^2} \left(r_0^2 \omega ^2+2 W^2\right) (r_0-x \cos (t \omega )-y \sin (t \omega )) 
 -4 c^5 m-2 c^4 \hbar  f''(z)+  \nonumber \\ 
 & 2 c^3 m f'(z)^2 \left(r_0^2 \omega ^2+2 W^2\right) + 2 c^3 m r_0^2 \omega ^2+c^2 r_0^2 \omega ^2 \hbar  f''(z)+r_0^2 \omega ^3 \hbar  \sqrt{c^2-r_0^2 \omega ^2} \sqrt{1-f'(z)^2}  -4 c^4 m \sqrt{c^2-r_0^2 \omega ^2}
 \nonumber \\& -2 c^3 \hbar  \sqrt{c^2-r_0^2 \omega ^2} f''(z)+c r_0^2 \omega ^3 \hbar  \sqrt{1-f'(z)^2}
 \bigg{)}, \\
 eA_1=& \frac{1}{4 c \sqrt{c^2-r_0^2 \omega ^2} \sqrt{1-f'(z)^2}} \bigg{(}
     -eB_0 \sqrt{1-f'(z)^2} \left(-2 c^2 r_0 \sin (t \omega )+2 c^2 y+r_0^2 x \omega ^2 \sin (2 t \omega )-r_0^2 y \omega ^2 \cos (2 t \omega ) 
      -r_0^2 y \omega ^2\right)  \nonumber \\& -4 c^2 m r_0 \omega  f'(z)^2 \sin (t \omega )+2 r_0 \omega  \sin (t \omega ) \left(2 c^2 m+c \hbar  f''(z)-\omega  \hbar  \sqrt{1-f'(z)^2}\right)
 \bigg{)}, \\
 eA_2 =&  \frac{1}{4 c \sqrt{c^2-r_0^2 \omega ^2} \sqrt{1-f'(z)^2}} \bigg{(}
   eB_0 \sqrt{1-f'(z)^2} \left(-2 c^2 r_0 \cos (t \omega )+2 c^2 x+r^2 x \omega ^2 \cos (2 t \omega )+r_0^2 y \omega ^2 \sin (2 t \omega )-r_0^2 x \omega ^2\right)
    \nonumber \\ & +4 c^2 m r \omega  f'(z)^2 \cos (t \omega )+2 r_0 \omega  \cos (t \omega ) \left(-2 c^2 m-c \hbar  f''(z)+\omega  \hbar  \sqrt{1-f'(z)^2}\right)
  \bigg{)}, \\
  eA_3 =& 0 ,
\end{align*}
with $W = c \sqrt{c^2-r_0^2 \omega ^2}+c^2-r_0^2 \omega ^2 $.

\begin{figure}
  \centering
      \includegraphics[width=0.5\hsize]{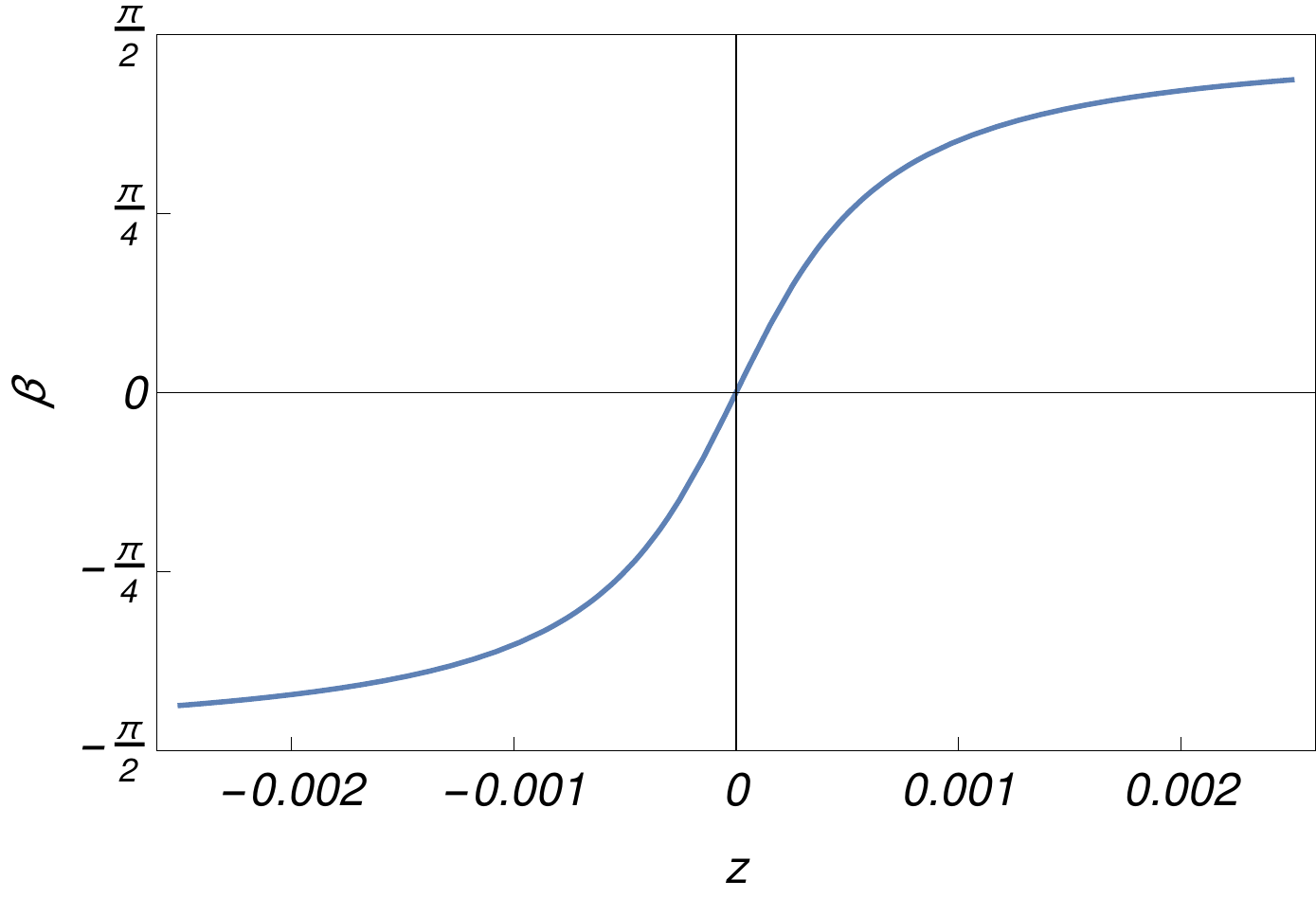}
      \caption{ Ivon-Takabayashi angle $\beta = \arcsin( f'(z) ) $  for $f(z) = \sqrt{ z^2 + \xi^2}$ employed in the main text
      ($\xi=5$pm). The larger the value of $\beta$, the larger is the contribution of negative energy particles.  }
  \label{fig:Takabayashi}
\end{figure}

\section{VI: Additional example with a scalar potential interaction}

\begin{figure}
  \centering
      \includegraphics[width=0.5\hsize]{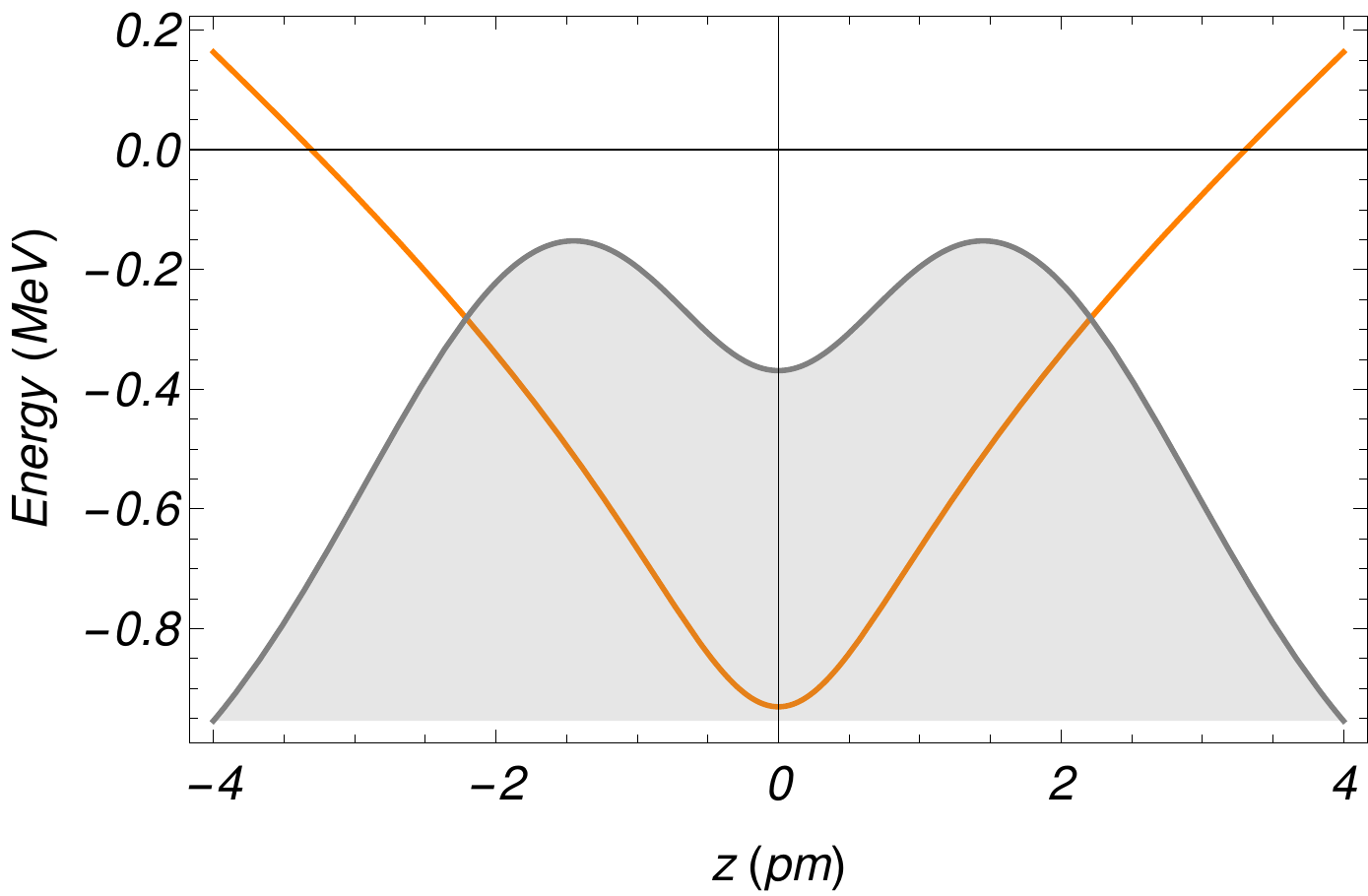}
      \caption{An exact stationary solution (gray shaded region) of the Dirac equation with 
      a scalar potential (orange line) and no electromagnetic fields.  
      The paramaters in Eqs. (\ref{Psi-Scalar}) and (\ref{V-Scalar}) 
      are $\xi=1 pm$, and $\epsilon=0.2 MeV$.  }
\label{smoothAbsPotential}
\end{figure}

RDI is not restricted to the construction of electromagnetic fields as shown in the main text.
In particular, we saw how an unbounded scalar potential can hold a stationary state.
In this section we show another example for a scalar interaction bounded from below with the absence of electromagnetic interactions. 
Supplying the stationary state with the predefined Yvon-Takabayashi angle,   
\begin{align} 
  \Psi =& 
   e^{ i \arctan[ z/\xi ]/2  }  e^{ - mc f(z)/\hbar  } 
   e^{ - i \sigma_3 \epsilon t/\hbar } , 
\label{Psi-Scalar}
\end{align}
the requirement of $A_0=0$ specifies both the scalar potential and an arbitrary function $f(z)$
\begin{align}
\label{V-Scalar}
V(z) &= -m c^2 +\frac{\epsilon}{\xi} \sqrt{z^2+\xi^2} - \frac{\hbar c}{ 2\sqrt{z^2+\xi^2}},\\
f(z) &= \frac{\epsilon(  z^2 + \xi^2 ) }{ 2 mc^2 \xi } z^2 - \frac{\hbar}{4mc} \log( z^2 + \xi^2 ) . \notag
\end{align}
A plot of the potential (\ref{V-Scalar}) along with the state (\ref{Psi-Scalar}) is shown in 
Fig. \ref{smoothAbsPotential}.

\bibliography{bib-relativity}

\end{document}